\documentclass[preprint]{ptephy_v1}

\preprintnumber{000-001} 
\usepackage{hyperref}
\usepackage{amsmath}
\usepackage{mathrsfs}
\usepackage{color}
\usepackage{comment}
\usepackage{fancyvrb}
\usepackage{graphicx}
\usepackage{amsmath}
\usepackage{amssymb}
\usepackage{bm}
\usepackage{amsfonts}
\usepackage{times}
\usepackage{epsfig}
\usepackage{dcolumn}
\usepackage{bm}
\usepackage{color}
\usepackage{longtable}
\usepackage{array}
\usepackage{multirow}
\usepackage{verbatim}
\usepackage[utf8]{inputenc}
\usepackage{scalerel}
\usepackage{graphicx}
\usepackage[normalem]{ulem}
\usepackage{subcaption}
\usepackage{lineno}
\usepackage{physics}
\usepackage{wrapfig}
\usepackage{float}

\VerbatimFootnotes

\newcommand{\ee}{e^+ e^-}

\newcommand{\pipi}{\pi^+\pi^-}

\newcommand{\jpsi}{J/\psi}
\newcommand{\pip}{\pi^+}
\newcommand{\pim}{\pi^-}
\newcommand{\piz}{\pi^0}
\newcommand{\KS}{K_{S}}
\newcommand{\KL}{K_{L}}
\newcommand{\Kone}{K_{1}}
\newcommand{\Ktwo}{K_{2}}
\newcommand{\Kp}{K^{+}}

\newcommand{\Kz}{K^{0}}
\newcommand{\Kzbar}{\bar{K}^{0}}

\newcommand{\Bz}{B^{0}}
\newcommand{\Bzbar}{\bar{B}^{0}}

\newcommand{\BH}{B_H}
\newcommand{\BL}{B_L}
\newcommand{\Btag}{B_{\rm tag}}
\newcommand{\BCP}{B_{\CPa}}

\newcommand{\Dm}{D^{-}}

\newcommand{\ra}{\rightarrow}

\newcommand{\lra}{\leftrightarrow}

\newcommand{\rt}{\rightarrow}

\newcommand{\BR}{{\cal B}}

\newcommand{\C}{${\mathcal C}$}
\newcommand{\Par}{${\mathcal P}$}
\newcommand{\CP}{${\mathcal C}{\mathcal P}$}
\newcommand{\CPa}{{\mathcal C}{\mathcal P}}
\newcommand{\T}{${\mathcal T}$}

\newcommand{\Bty}{${\mathcal B}$}

\newcommand{\DelS}{\Delta{\mathcal S}}

\newcommand{\Lum}{{\mathcal L}}

\newcommand{\eps}{\varepsilon}
\newcommand{\epsp}{\varepsilon^{\prime}}

\newcommand{\Ups}{{\Upsilon}}

\newcommand{\Ecm}{E_{\rm cm}}

\newcommand{\Vud}{V_{ud}}

\newcommand{\Vub}{V_{ub}}
\newcommand{\Vcd}{V_{cd}}

\newcommand{\Vcb}{V_{cb}}
\newcommand{\Vtd}{V_{cd}}

\begin{document}

\title{The Curious Early History of CKM  Matrix\\
  -{\it miracles happen!}-}


\author{Stephen Lars Olsen}
\affil{Institute for Basic Science, Daejeon South Korea \\
CAU High Energy Physics Center, Chung-Ang University, Seoul South Korea\email{solsensnu@gmail.com} }


\begin{abstract}%

  \noindent
  The 1973 Kobayashi Maskawa paper proposed a compelling link between Cabibbo’s flavor-mixing scheme and \C\Par
  violation but, since it required the existence of six quarks at a time when the physics community was happy
  with only three, it received zero attention.  However, two years after the paper appeared---at which time
  it had received a grand total of two citations---the charmed quark was discovered and it finally got some
  notice and acceptance.  After this stumbling start, it subsequently emerged as the focal point of an enormous
  amount of experimental and theoretical research activity. In an invited talk at a KEK symposium to celebrate the
  50$^{\rm th}$ anniversary of the KM paper, I reviewed some of the less well known circumstances that occurred
  in the years preceding and following the paper's appearance.

  \vspace{1mm}
  \noindent
  Some spoilers:
  
 \vspace{1mm}
 \noindent
---~Kobayashi and Maskawa (and a number of other Japanese physicists) were convinced about the existence of the
 charmed quark nearly three years before its ``discovery'' at Brookhaven and SLAC.

 \vspace{1mm}
 \noindent
---~The matrix provided in their seminal 1973 paper was mathematically incorrect. Another version
that was in common use for the following twelve years was technically correct, but not really a
rotation matrix.

  \vspace{1mm}
  \noindent
---~The CKM matrix \C\Par~phase was only measurable because of the very specific hierarchy
  of the flavor mixing angles and meson masses.

 \vspace{1mm}
   \noindent
---~Similarly, the neutrino mixing discovery, and the PMNS-matrix measurability were only possible because of
   favorable values of the neutrino mass differences and mixing angles.

\vspace{2.5mm}
   \noindent
   In addition I include some speculations about what may be in store for the future. 

\end{abstract}

\subjectindex{Quark flavor mixing, Neutrino flavor mixing, C\Par~violation}

\maketitle

\section{Introduction}

\noindent
The challenge of reviewing a subject that is fifty years old to a community of experts is to find something to
say that isn't already well known to everyone in the audience. However, this obvious truth didn't occur to me
when I was invited by the organizers to speak at the KEK special symposium to celebrate the fiftieth anniversary
of the Kobayashi-Maskawa six-quark model. An invitation that, in a reckless capitulation to my vanity, I immediately
accepted. Upon subsequent reflection, I realized my dilemma: there was precious little that I could say about the
hundreds of CKM-related published Belle results---which I expect the organizers had in mind when they offered
this invitation---that wasn't already very familiar to the symposium participants. So, instead, I decided to exploit the
one advantage I might have over most other participants, and that was that I would be the oldest, or least one of
the oldest, person in attendance and reminisce about the early days of the KM era, including some of its pre-history.
So, with the forewarning that all historical accounts suffer from mistakes and oversimplifications, and are varnished
to match the preconceptions and prejudices of the chronicler, here goes:

\section{Prehistory: Cabibbo flavor-mixing and the discovery of \C\Par~violation}

\noindent
The  prehistory started sixty years ago during the 1963-64 academic year\footnote{This happened
       to coincide with my first year as a graduate student at the University of Wisconsin.}
when there were three major discoveries that all played a major roles in the Kobayashi-Maskawa story:
flavor-mixing, quarks, and the observation of \C\Par~violation in $\KL$$\rt$$\pipi$ decays.

\subsection{Cabibbo flavor mixing}
\begin{figure}[!tbp]
  \centering
     \includegraphics[width=0.85\textwidth]{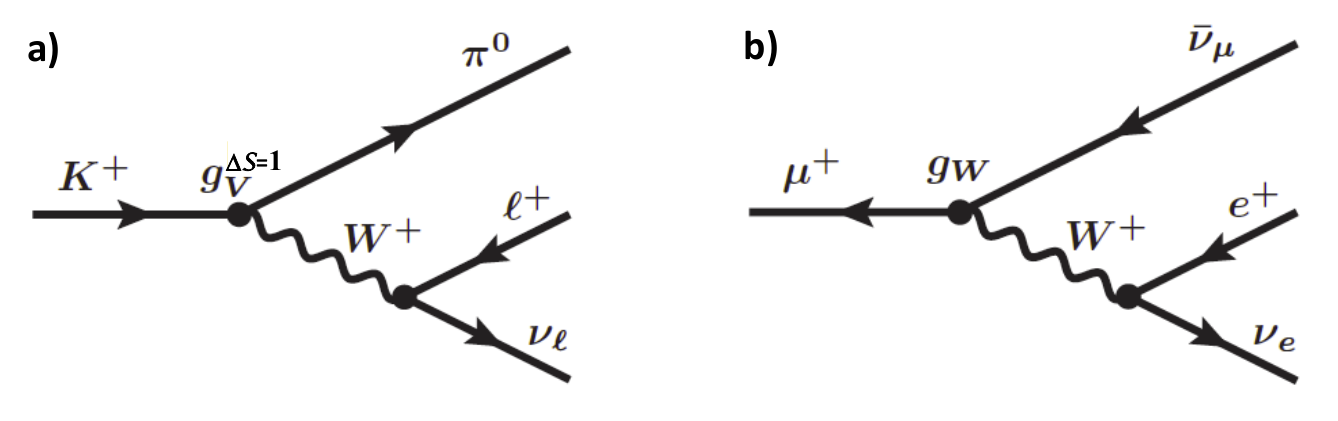}
     \caption{\footnotesize {\bf a)} {\it (upper)} The lowest order Feynman diagram for  $K^+$$\rt$$\piz\mu^+\nu$.    
       {\bf b)} The lowest-order Feynman diagram for $\mu^+$$\ra$$e^+{\nu}_e\bar{\nu}_{\mu}$.
    }
    \label{fig:Cabibbo-angle}
\end{figure}

\noindent
In their classic paper that identified the $V$$-$$A$ coupling of the  weak interaction~\cite{Feynman:1958ty},
Feynman and Gell-Mann proposed that the weak interaction was a current-current interaction where the
hadron current has the form
     \begin{equation}
       \label{eqn:QWI-current}
       J_{\mu}=g\big[\alpha \big(V_\mu^{\DelS=0}-A_\mu^{\DelS=0}\big)+
                      \beta\big(V_\mu^{\DelS=1}-A_\mu^{\DelS=1}\big)\big],
     \end{equation}
     where $g$ is a coupling constant,
     $V^{\DelS=0}_{\mu}$~and~$A^{\DelS=0}_{\mu}$ are the vector and axial vector currents for strangeness
     conserving processes and  $V^{\DelS=1}_{\mu}$~and~$A^{\DelS=1}_{\mu}$ are corresponding currents for
     $\DelS=\pm1$ transitions. 
      They also made two additional
      conjectures. One was {\it universality}, the notion that the currents for the  $\DelS=0$ and $\DelS=\pm 1$
      hadronic transitions and the
\begin{equation}
  g_W\big(\bar{\nu}_e\gamma_\mu(1-\gamma_5)e^-\big)~~~~{\rm and}
          ~~~~g_W\big(\bar{\nu}_\mu\gamma_\mu(1-\gamma_5)\mu^-\big)
\end{equation}
     lepton currents all have a common coupling strength, {\it i.e.}, $g=g_W$, and
     $\alpha$\,=\,$\beta$\,=\,1 in eqn.~\ref{eqn:QWI-current}, where $g_W$ is related to the square root
     of the Fermi constant $G_F$ by
     \begin{equation}
       G_F=\frac{\sqrt{2}}{8}\bigg(\frac{g_W}{M_W}\bigg)^2.
     \end{equation}
     \noindent
     The other one was the so-called {\it Conserved Vector Current} (CVC) hypothesis that says that the hadronic
     matrix elements for the vector component of the weak interaction current are the same as those for the
     electromagnetic interactions. This has the consequence that vector form-factors for weak decays of hadrons
     at zero squared momentum-transfers are unity, $f_V(q^2$=$0)$\,=\,1. These two conjectures translated into a
     prediction that the coupling strength extracted from the vector-mediated semileptonic process
     $K^+$$\ra$$\piz e^+\nu_e$, {\it i.e.}, $g_V^{\DelS=1}$ shown in Fig.\ref{fig:Cabibbo-angle}a) should be 
     the same as $g_W$ in $\mu^+$$\rt$$e^+\nu_e\bar{\nu}_\mu$.
      
     In a paper that appeared in June 1963~\cite{Cabibbo:1963yz}, Cabibbo pointed out that Feynman-Gell-Mann
     universality conjecture failed miserably. His comparison of experimental measurements of the partial width
     for the $\DelS=1$ vector weak-interaction process $\Kp$$\rt$$\piz \ell^+\nu$~\cite{Roe:1961zz} to the well
     known width for muon decay found
     
     \begin{equation}
      \frac{g_V^{\DelS=1}}{g_W}\approx 0.26
     \end{equation}
     \noindent
 and about a factor of four below expectations.  He also found a similar deviation from universality in the ratio
 of the axial-vector-mediated partial decay widths $\Gamma(\Kp$$\rt$$\mu^+\nu)$/$\Gamma(\pip$$\rt$$\mu^+\nu)$:
      \begin{equation}
        \frac{g_A^{\DelS=1}}{g_A^{\DelS=0}}\approx  0.26. 
      \end{equation}
      (Although the axial-vector currents are not ``protected'' by CVC, corrections to them were expected to be
      small~\cite{GellMann:1960np}, and  certainly not large enough to account for a factor
      of four.)

Cabibbo proposed modifying the Feynman-Gell-Mann $\alpha$\,=\,$\beta$\,=\,1 conjecture to $\alpha^2$+$\beta^2$=\,1,
      in which case
     \begin{equation}
       g_V^{\DelS=1} = \beta g_W ~~~~{\rm and}~~~~\frac{g_A^{\DelS=1}}{g_A^{\DelS=0}}=\frac{\beta}{\alpha},
      \end{equation}
 where $\beta\approx 0.25$ could accommodate the abovementioned experimental results. In his paper, Cabibbo proposed
     his eponymous angle $\theta_C$, which he estimated to $\theta_C\approx 14.9^\circ$, as a convenient way to
     express two parameters $\alpha$~and~$\beta$ that were subject to the constraint $\alpha^2$+$\beta^2$=1, and he
     didn't mention  anything about rotations. The earliest experiments that addressed
     Cabibbo's hypothesis~\cite{Eisele:1969gw} were focused on testing the validity of Cabibbo's relation,
     $\alpha^2$\,+\,$\beta^2$\,=\,1.

     The notion that this might represent a rotation didn't become apparent until the 1970 GIM
     paper~\cite{Glashow:1970gm} that
     proposed the $c$-quark as a way to suppress flavor-changing neutral currents. If one accepts the
     existence of two quark doublets,

     \begin{equation}
      \begin{pmatrix}u\\d\end{pmatrix}{\displaystyle s}~\Longrightarrow~       
                                   \begin{pmatrix}u\\d\end{pmatrix}\begin{pmatrix}c\\s\end{pmatrix},
     \end{equation}
     the Cabibbo $d$-$s$ mixed quark state $d^{\prime}$\,=\,$d\cos\theta_C$\,+\,$s\sin\theta_C$
     is produced by the application of a  2x2  unitary  rotation  matrix:
     \begin{equation}
     \label{eqn:2by2-flavor-mixing-matrix}
       \begin{pmatrix}d^{\prime}\\s^{\prime}\end{pmatrix}=
         \begin{pmatrix}\cos\theta_C & \sin\theta_C\\-\sin\theta_C & \cos\theta_C\end{pmatrix}
            \begin{pmatrix}d\\s\end{pmatrix}=
               \begin{pmatrix} d\cos\theta_C +s\sin\theta_C\\-d\sin\theta_C  +s\cos\theta_C\end{pmatrix},
     \end{equation}
     and has an orthogonal partner, $s$\,=$-d\sin\theta_C$\,+\,$s\cos\theta_C$. In this formulation,
     it is apparent that
Cabibbo's form of weak universality is  the same as Feynman-Gell-Mann universality applied to the rotated
$d^\prime$\,and\,$s^\prime$ quarks.\footnote{In addition to suppressing $\DelS$\,=$\pm 1$
  weak interaction couplings relative to that for muon decay by a factor of
  $\sin\theta_C$\,=\,$0.2245$, Cabibbo's weak universality predicts that $\DelS$\,=\,0
  couplings are suppressed by a factor of $\cos\theta_C$\,=\,$0.974$. In fact, nuclear
  physicists had known since 1955 that the half-life for $^{14}{\rm O}$$\rt$$^{14}{\rm N}\beta^+\nu$,
  a vector-mediated $0^+$$\ra$$0^+$ nuclear $\beta^+$ decay transition, was $\sim$3\%~longer
  than the value that was  predicted using the $g_W$ value determined from muon
  decay~\cite{Gerhart:1954zz,Hendrie:1961zz}. In 1960, three years before Cabibbo's
  paper, this discrepancy was noted in the introductory remarks of a Nuovo Cimento
  article on the axial-vector current by Gell-Mann and Levy~\cite{GellMann:1960np},
  together with a footnote that suggested that this might be because the unitarity
  condition might be, in fact, $\alpha^2$+$\beta^2$=\,1, and not the
  $\alpha$\,=\,$\beta$\,=\,1~condition that was conjectured in the Feyman-Gell-Mann
  $V$-$A$ paper.  The footnote includes a estimate on the mixing that translates into
  $\theta$\,$\approx$\,14$^\circ$, consistent with---and three years before---Cabibbo's
  estimate for $\theta_C$ based on $\DelS$\,=\,1~transitions.  This may explain why
  K\,and\,M, but not C, were awarded the Nobel prize in 2008.}

\subsection{Gell-Mann Zweig quarks}
\noindent
     During this same year Gell-Mann~\cite{GellMann:1964nj} and Zweig~\cite{Zweig:1981pd} proposed the quark model
     in which hadrons were comprised of fractionally charge fermionic constituents (Zweig called then ``aces'').
     Gell-Mann's paper was published in January 1964; Zweig's paper was never published.\footnote{The
       story here is that the head of the CERN theory group in 1964, when Zweig was there on
       a visiting appointment, thought Zweig's proposed fractionally charged particles
       was a crackpot idea and refused to provide him with the clerical and drafting
       support that was needed to prepare a journal-worthy manuscript in the
       pre-Latex\,\&\,computer-graphics era. Gell-Mann won the 1969 Nobel physics prize
       and by 1976, when the head of the theory group became the Director-General of CERN,
       Zweig was doing biological research and no longer involved in particle physics.}
     With rotated quarks, the short-distance weak interaction hadronic currents are the same as those for leptons:
     \begin{eqnarray}
       \label{eqn:wi-quark-current}
       J_{\mu}^q &=& g_W(\bar{u}\gamma_\mu(1-\gamma^5)d^\prime)~+~g_W(\bar{c}\gamma_\mu(1-\gamma_5)s^\prime) \\
       \nonumber
                &=& g_W\sum_{i,j}(\bar{u}_i\gamma_\mu(1-\gamma^5)V_{ij}d_j),
     \end{eqnarray}
     where $(u_1,u_2)$\,=\,$(u,c)$~\&~$(d_1,d_2)$\,=\,$(d,s)$, and~$V_{ij}$ is the
     eqn.~\ref{eqn:2by2-flavor-mixing-matrix} quark mixing matrix. The long distance quark-to-hadron
     processes are described by form factors.

\subsection{Discovery of \C\Par~violation}
\begin{figure}[htb!]
\centering
  \includegraphics[width=1.0\textwidth]{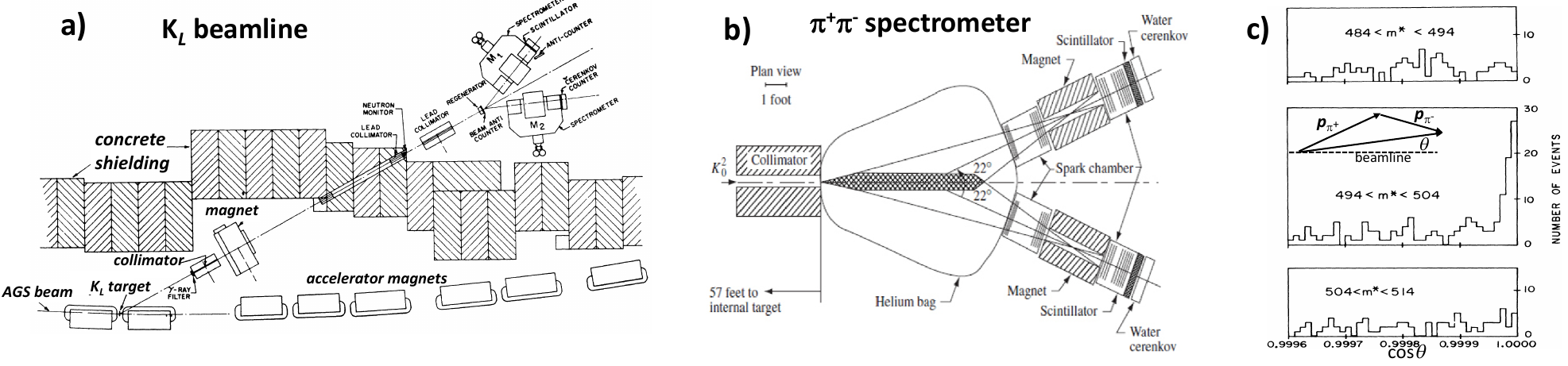}
  \caption{\footnotesize
    {\bf a)} The neutral $\KL$-beamline at the AGS that was used for the $\KL$$\ra$$\pipi$ search
    experiment.
    {\bf b)} The two-arm $\pipi$ spectrometer consisted of a helium-filled decay volume
    followed by optical tracking spark chambers before and after momentum analyzing magnets
    {\bf c)} The distribution of events {\it versus} the cosine of the angle between the
    direction of the two-track momentum sum and the beamline. The upper, middle and lower
    panels are for events with two-track invariant masses that are below, centered on, and
    above $m_{\KL}$, respectively.
    }
    \label{fig:cronin-fitch-expt}
 \end{figure}

\noindent
The  Christenson, Cronin, Fitch and Turley discovery of the \C\Par~violating decay mode $\KL$$\rt$\,$\pipi$ was
reported in the summer of 1964~\cite{Christenson:1964fg}.  This was a relatively low priority experiment that
was not aimed at investigating \C\Par~violation but, instead, was designed to investigate some anomalies in coherent
$\Ktwo$$\rt$$\Kone$ regeneration measurements that had been reported during the previous
year~\cite{Leipuner:1963zz}. It failed to qualify for a spot in
the main experimental hall of the then, almost new, AGS synchrotron that was occupied by spectrometers specialized
for total cross section determinations, and $\pi$, $K$, $\bar{p}$ and $\mu$-proton elastic scattering
measurements. Instead, the experimental apparatus was located in a relatively inaccessible area inside
the AGS magnet ring that the laboratory technical staff referred to as ``Inner Mongolia,''\footnote{In
  the 1960s diplomatic relations between the U.S. and China were non-existent, and
  mainland China, including Inner Mongolia, was considered by most Americans to be about as accessible as the
  far side of the Moon.}
in a neutral particle line that was essentially a hole in the AGS shielding wall that was pointed at a target
located in the accelerator's vacuum chamber, as illustrated in Fig.~\ref{fig:cronin-fitch-expt}a. The high flux
of $\gamma$-rays emerging from the target were attenuated by a 3.8~cm-thick lead block followed by a collimator
and a bending magnet that swept charged particles out of the beam aperture. A double-arm spectrometer consisting
of tracking spark chambers before and after two vertically bending magnets measured the directions and momenta
of charged particles that were produced by $\KL$ meson decays that occurred in a 2~m-long decay volume that was a
plastic bag filled with atmospheric pressure helium---a low-budget approximation of a vacuum chamber---as shown
in Fig.~\ref{fig:cronin-fitch-expt}b.

Most of the detected events were due to \CP-allowed $\KL$$\ra$$\pipi\piz$ decays and
$\KL\ra\pi^{\pm}\ell^{\mp}\nu$ ($\ell$\,=\,$e,\mu$) {\it semileptonic} decays. In these decays, the $\piz$ or
$\nu$ was not detected and, as a result, the invariant mass of the two detected charged particles
was not, in general, equal to the $\KL$-meson mass ($m_{\KL}$\,=\,$498$~MeV). For $\KL$$\ra$$\pipi\piz$ decays
where the $\piz$ is undetected, the $\pipi$ invariant mass is always below 363~MeV; for
$\KL$$\rt$$\pi^{\pm}\ell^{\mp}\nu$, where the $\nu$ is missed and the $\ell^{\mp}$ track is assigned a pion
mass, the two charged track invariant mass distribution ranges from 280\,to\,546~MeV, with no peak near
$m_{\KL}$. Although the energies of the decaying $\KL$ mesons were not known, their three-momentum directions
were confined to be within an rms spread of $\pm 3.4$~mrad ($\pm 0.2^\circ$) around the beamline. A consequence
of the missed particle in the three-body decay channels was that the vector sums of the two charge track's momentum
vectors did not usually point along the well defined $\KL$ beamline.

Thus, the experimental signature for $\KL$$\rt$$\pipi$ decays was a pair of oppositely charged tracks that, when
assigned pion masses, had an invariant mass that was within $\pm 5$~MeV of $m_{\KL}$ and with a summed vector
momentum that is directed along the $\KL$ beam direction. Results for these two quantities are shown in
Fig.~\ref{fig:cronin-fitch-expt}c, where the horizontal axis is the cosine of the angle between
the $\vec{p}_{\pip}$\,+\,$\vec{p}_{\pim}$ direction and the $\KL$ beamline, and the upper, central and lower
panels show the experimental distributions for $M(\pipi)$ below, centered on, and above $m_{\KL}$,
respectively. In the central panel there is a pronounced peak totally contained within
$\cos\theta$\,$>$\,$ 0.99996$ ($\theta$\,$<$\,$9$~mrad), a feature that is absent in the distributions
for $M(\pipi)$ below or above $m_{\KL}$ shown in the upper and lower panels. The 
ratio of the branching fractions for $\KL$$\ra$$\pipi$ to the sum of all (\C\Par-conserving) decays to
charged particles  was $(2.0\pm 0.4)$\,$\times$\,$10^{-3}$.

The signal peak in the central panel of Fig.~\ref{fig:cronin-fitch-expt}c contained an excess of
$45\pm 10$ events, but these were not all $\KL$$\rt$$\pipi$ events.  About ten of them were due to
the coherent $\KL$$\rt$$\KS$$\rt$$\pipi$ regeneration process on the helium nuclei in the gas bag
decay region, and were indistinguishable from the $\KL$$\rt$$\pipi$ signal events. Nature was kind.
If the branching fraction had been much smaller or the regeneration cross section were higher, the
interpretation of the observed signal peak would have been ambiguous. As mentioned above, this
was a low-priority experiment. If it had ended up by simply setting an upper limit on the
$\KL$$\rt$$\pipi$ branching fraction, who knows when, if ever, a follow-up experiment with higher
sensitivity would have occurred.

\section{The Kobayashi-Maskawa paper}

\noindent
The famous Kobayashi-Maskawa paper~\cite{Kobayashi:1973fv} was written in mid-1972,
and published in the February 1973 issue of the Japanese journal {\it Progress of Theoretical Physics},
where it was basically ignored; during the following two~and~a~half~years, it received all of two citations. 
The paper's title is \CP~{\it Violation in the Renormalizable Theory of Weak Interaction}, where
the Renormalizable Theory of Weak Interaction is the term they used for what we now call the
Standard Model.

Simply put, a \C\Par~violation means that the amplitudes for a processes that involve initial-state
particles converting to final-particles and its corresponding antiparticle equivalent are not the same,
{\it e.g.},
\begin{equation}
  {\mathcal M}(a\rt bc)=\bra{bc}H_w\ket{a} \neq
  \bar{\mathcal M}(\bar{a}\rt \bar{b}\bar{c})=\bra{\bar{b}\bar{c}}H^{\dag}_w\ket{\bar{a}}.
\end{equation}
But the hermiticity of the Hamiltonian requires that the squares of the amplitudes are equal:
\begin{equation}
  |{\mathcal M}(a\rt bc)|^2=|\bra{bc}H_w\ket{a}|^2 =
  |\bar{\mathcal M}(\bar{a}\rt \bar{b}\bar{c})|^2=|\bra{\bar{b}\bar{c}}H^{\dag}_w\ket{\bar{a}}|^2,
\end{equation}
and the only way these two conditions can be satisfied is if ${\mathcal M}$ and $\bar{\mathcal M}$
differ by a phase, {\it i.e.},
\begin{equation}
  {\mathcal M}=|{\mathcal M}|e^{i\delta_{\CPa}}~~~~{\rm and}
          ~~~~\bar{\mathcal M}=|{\mathcal M}|e^{-i\delta_{\CPa}}.
\end{equation}
So, to incorporate \C\Par~violation into the Standard Model, all you have to do is find a way to insert a complex
phase in it somewhere, which, at first glance, wouldn't seem to  be so difficult.  However this \C\Par-violating
phase is special and, unlike all other phases that show up quantum theories, and including it into the the
theory is not at all trivial.

There are countless phases that occur in quantum mechanics; both the Schr{\" o}dinger and Dirac
equations have an imaginary coefficient and their solutions are complex wave functions. But
all of these phases, with, so far, only one exception, have the same sign for particles
and antiparticles. Only a \C\Par-violating~phase has opposite signs for particles and antiparticles.

The KM paper examines various possible ways that a complex \C\Par~phase might be incorporated
into the Standard Model.  In the following I discuss the first five pages and the last
page separately.

\subsection{The KM paper: pages 1$\rt$5}
\noindent
In the first five pages, various possibilities were examined and the authors concluded that
``no realistic models of \C\Par-violation exist in the {\it quartet scheme} without
introducing any other new fields.''  Here, by the ``quartet scheme'' they meant the four-quark
model that included the charmed quark. Note that this was written in 1971, nearly three
before the $c$-quark discovery in the ``November 1974 revolution.''  This, and their page-five conclusion
that no realistic model for \C\Par-violation exists with four quarks raise two questions:

\vspace{1mm}
\noindent
{\it i}) Why were Kobayashi and Maskawa so sure of the existence of the $c$-quark at
such an early data?

\vspace{1mm}
\noindent
{\it ii}) Why can't a \C\Par-violating phase be introduced together with the Cabibbo angle
into the eqn.~\ref{eqn:2by2-flavor-mixing-matrix} 2$\times$2 quark-flavor mixing matrix?

\subsubsection{The discovery of charm: Japanese version}
\begin{figure}[htb!]
\centering
  \includegraphics[width=0.9\textwidth]{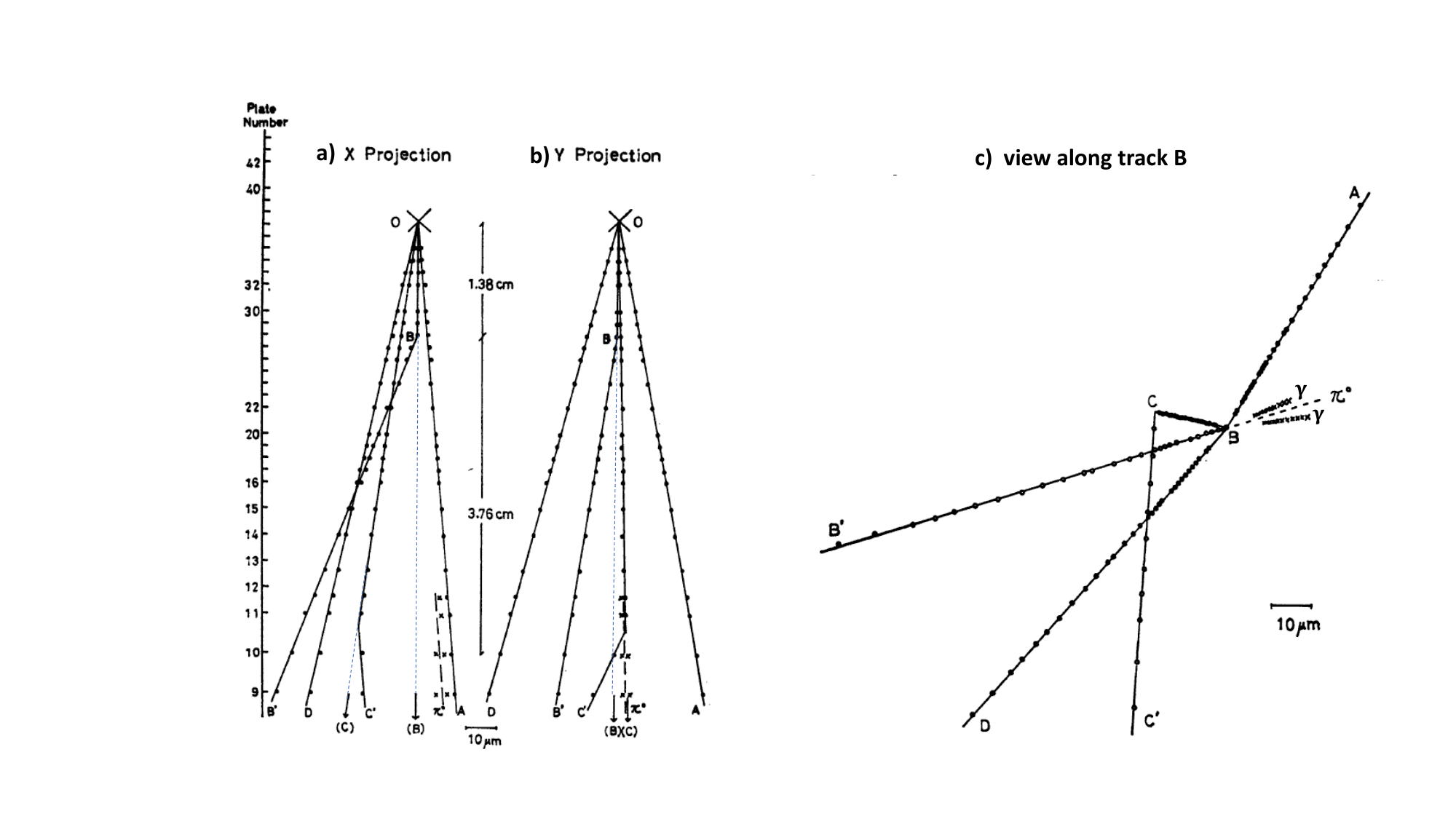}
  \caption{\footnotesize
    The {\bf a)} $X$ and {\bf b)} $Y$ projections of the Niu event. Here the tracks labeled B
    and C have kinks at depth of 1.38~cm and 5.14~cm, respectively, that are evident in both views.
    {\bf c)} The same event viewed along track B, where the direction of a high energy $\piz$,
    inferred from two detected $\gamma$-rays, is very nearly opposite the direction of B', the
    daughter track that emerges from the track B kink. (The figures are taken from ref.~\cite{Niu:1971xu}.)
  }
    \label{fig:niu-event}
 \end{figure}

\noindent
In 1970, a small team of experimenters in Japan led by Kiyoshi Niu, exposed a stack of photographic
emulsions to cosmic rays in a high altitude commercial cargo airliner~\cite{Niu:1971xu}. Upon subsequent
inspection they found a remarkable event, shown in Fig.~\ref{fig:niu-event}, in which an ultra-high
energy (multi-TeV) cosmic ray particle interaction
produced four charged tracks and two very high energy, closely spaced $\gamma$-rays that, when
attributed to a $\piz\ra\gamma\gamma$ decay, had a total energy of $3.2\pm 0.4$~TeV. Two of the charged
tracks, labeled B\,\& \,C in the figure, have kinks within $\sim$5~cm of the production point that are
quite distinct in both the $X$ and $Y$ projections shown in Figs.~\ref{fig:niu-event}a\,\&\,b,
indicating that they decayed to charged daughters (tracks B' \& C'). When the event is viewed along the
flight direction of track B (Fig.~\ref{fig:niu-event}c), its daughter charged track (B') and the high
energy $\piz$ are very close to being back-to-back.  The transverse momentum of the $\piz$ relative to the
direction of track B was $627\pm 90$~MeV, and much higher than was possible for the decay of any known
particle at that time. With the $\piz$ setting the energy scale and  assumptions of two-body decays at each
kink: $\pi^{\pm}\piz$ for B$\rt$B' and $\piz p$ for C$\rt$C' where the secondary $\piz$ is missed, transverse
momentum balance was used to estimate the masses and lifetimes of B and C:  
\begin{center}
\begin{tabular}{c|c|c}
\hline
Assumed            &   Mass  & lifetime\\
decay mode         &  (GeV)  &  (sec)  \\
\hline
B$\rt$$\pip\piz$  &  1.78   &  2.2$\times 10^{-14}$\\
C$\rt$$(\piz) p$    &  2.95   &  3.6$\times 10^{-14}$,\\
\hline
\end{tabular}
\end{center} 
The estimated B mass and the proper time intervals are consistent with the GIM estimates of
$\sim$2~GeV for the charmed-quark mass (and in reasonable agreement with what the now very well
determined $\Dm$ mass (1.869~GeV) and $\Lambda_c$ mass (2.286~GeV)). The lifetimes were much shorter than
that of any known weakly decaying particle as well as the ${\mathcal O}(10^{-13\,}{\rm s}$)  estimate
that was given in the GIM paper. But the latter fact is perhaps not too surprising since emulsion measurements
are biased towards shorter lifetimes.
For these reasons, Nagoya theorist Shuzo Ogawa interpreted Niu's event as being the associated production
of an anticharmed meson charmed-baryon pair and their subsequent decays. Although whether or not Niu's event and
Ogawa's interpretation amounted to a Nobel-prize-worthy claim of a discovery might be a subject of dispute,
what matters for our story here is that many people in the Japanese theoretical physics community, especially
those in  Nagoya that included Kobayashi and Maskawa, were convinced that the charmed quark had been discovered,
and that four quarks existed in nature, a scenario that they called the ``quartet model.''

\subsubsection{A \C\Par~phase in the four-quark mixing matrix?}
\noindent
In general, a  2$\times$2 matrix like that in eqn.~\ref{eqn:2by2-flavor-mixing-matrix} has four complex
elements that correspond to eight distinct real numbers.  In the four-quark model, flavor mixing is
completely described by the single real number $\theta_C$, why can't one of the other seven numbers
be used to specify $\delta_{\CPa}$, a \C\Par-violating phase?

The flavor-mixing matrix describes a rotation and, thus, has to conserve probability. This means it
should be unitary: {\it i.e.} $\boldsymbol{VV^\dag}$=\,$\boldsymbol{\mathcal I}$, where
$\boldsymbol{\mathcal I}$ is the identity matrix:
\begin{equation}
  \begin{pmatrix}V_{ud} & V_{us}\\V_{cd} & V_{cs} \end{pmatrix}
      \times\begin{pmatrix}V^*_{ud} & V^*_{cd}\\V^*_{us} & V^*_{cs} \end{pmatrix}
        =\begin{pmatrix} 1 & 0 \\ 0 & 1 \end{pmatrix}.
\end{equation}
This corresponds to four relations
\begin{eqnarray}
  \label{eqn:1st-row-unitarity}
  |V_{ud}|^2+|V_{us}|^2=1 ~~~~~&{\rm and}&~~~~~|V_{cd}|^2+|V_{cs}|^2=1\\
  V_{ud}V^*_{cd}=-V_{us}V^*_{cs}~~~~~&{\rm and}&~~~~~V_{cd}V^*_{ud}=-V_{cs}V^*_{us},
\end{eqnarray}
that reduces the number of independent parameters from eight to four.

In the weak interaction quark currents (eqn.~\ref{eqn:wi-quark-current}), the total number of quarks is conserved:
$q_j$, which annihilates a $q_j$-quark, is always accompanied by $\bar{q}_i$, which creates a $q_i$-quark. The
theory has a subtle property: if each quark field is multiplied by an arbitrary phase factor,
\begin{equation}
   d_j\longrightarrow e^{i\phi_j}d_j~~~~{\rm and}~~~~~\bar{u}_i\longrightarrow e^{-i\phi_i}\bar{u}_i,
\end{equation}
and the interactions are modified by the same phases,
\begin{equation}
  V\longrightarrow
  \begin{pmatrix}
    e^{i\phi_u}   &      0       \\
    0           &  e^{i\phi_c}  
  \end{pmatrix}
  \begin{pmatrix}
    V_{ud}       &     V_{us}   \\
    V_{cd}       &     V_{cs}  
   \end{pmatrix}
  \begin{pmatrix}
    e^{-i\phi_d}   &      0      \\
    0           &  e^{-i\phi_s}
   \end{pmatrix},
\end{equation}
there is no net effect on on the $J_{\mu}^q$ current: 
\begin{eqnarray}
  \nonumber
    (\bar{u}_i \gamma_{\mu}(1-\gamma_5)V_{ij}d_j)& \longrightarrow &
  (\bar{u}_i e^{-i\phi_i}\gamma_{\mu}(1-\gamma_5)e^{i(\phi_i-\phi_j)}V_{ij}e^{i\phi_j}d_j)\\
    &=& (\bar{u}_i \gamma_{\mu}(1-\gamma_5)V_{ij}d_j).
\end{eqnarray}
This process is called {\it rephasing} and the four phases can be expressed as three independent phase
differences plus one overall phase that has no effect on the physics. Thus, of the the eight real
numbers we started with, four are removed by the unitarity constraint, and three can have
any value with no net effect. Thus there is only one number
left to define the matrix, and that is needed for the rotation angle $\theta_C$.  There is no freedom to
add a \C\Par~phase and is what led Kobayashi and Maskawa to conclude that there was no way to
incorporate \C\Par~violation into the four quark model.

\subsection{The KM paper:  page 6}
\noindent
In his 2008 Nobel prize lecture~\cite{Maskawa:2009zza},  Maskawa recalled that he and Kobayashi
had completed the work that was covered in pages~1-5 of their paper and were disappointed with
their failure to find any way to incorporate a \C\Par~phase into the four quark model,
and were reconciled to the unhappy likelihood that they would have to publish the negative result.
Then, one evening, while---as is customary in Japan---he was taking his after-dinner bath, he mentally
went through the calculation described in the previous paragraph, except this time for a six-quark
scenario with a 3$\times$3 flavor-mixing rotation  matrix in three dimensions. In this case, there
are 9~complex elements that are described by 18~real numbers.  Of these, 9 are constrained by the
unitarity requirement, 5 are taken up by rephasing, and 3 are needed to specify the
3-dimensional rotation,\footnote{In general, for $N$ quark generations with
  an $N$$\times$$N$ mixing matrix, there are $N^2$ elements characterized by
  $2N^2$ real numbers. Of these $N^2$ are used for unitarity, $2N$-$1$ for
  rephasing and $N(N$-$1)/2$ are needed to define the rotation angles. The
  remaining degrees of freedom that can show up as \C\Par~phases are
  $(N$-$1)(N$-$2)/2$, which vanishes for $N$=$2$, but is~$1$for $N$=$3$
  (and would be~$3$~if $N$=$4$).}
and that left one number that remained available.

\vspace{2mm}
\noindent
{\large {\it Eureka!}}~~~~With six-quarks there is room for a \C\Par-violating phase!

\vspace{1.5mm}
\noindent
A sixth page was added to the manuscript that included the remarks:
\begin{quotation}
    {{Next we consider a 6-plet model, another interesting model of \C\Par~violation,... with
    a 3$\times$3 instead of 2$\times$2 unitary matrix. In this case we cannot absorb all
    phases of matrix elements into the phase convention and can take, for example, the
    following expression}}:
\end{quotation}
\begin{equation}
  \label{eqn:KM-matrix-orig}
   \begin{pmatrix}
  \cos\theta_1    &                 -\sin\theta_1\cos\theta_3             &      -\sin\theta_1\sin\theta_3 \\
  \sin\theta_1\cos\theta_2&\cos\theta_1\cos\theta_2\cos\theta_3-\sin\theta_2\sin\theta_3 e^{i\delta}&
                           \cos\theta_1\cos\theta_2\sin\theta_3 +\sin\theta_2\cos\theta_3 e^{i\delta} \\
\sin\theta_1\sin\theta_2&\cos\theta_1\sin\theta_2\cos\theta_3+\cos\theta_2\sin\theta_3 e^{i\delta}&
                           \cos\theta_1\sin\theta_2\sin\theta_3 -\cos\theta_2\sin\theta_3 e^{i\delta}
  \end{pmatrix}.
\end{equation}
\begin{quotation}
  {{Then we have \C\Par-violating effects ... that appear only in the $\DelS$$\neq$0
    non-leptonic processes and semi-leptonic decay of neutral strange mesons (we are not
    concerned with higher states with the new quantum number) ...}}
\end{quotation}
\noindent
(Here, $\theta_1$ is (approximately) the Cabibbo angle, $\theta_2$ is the mixing angle between the
2$^{\rm nd}$-~\&~ 3$^{\rm rd}$-generation quarks, $\theta_3$ mixes the 1$^{\rm st}$-~\&~3$^{\rm rd}$-generations,
and $\delta$ is the \C\Par-violating~phase.)
And that was it. Discussion about the six-quark model was confined to one paragraph that, together with
the expression for the matrix, occupied only about half of the page.  

Thus, Kobayashi and Maskawa discovered the way to incorporate a \C\Par~violating phase into the
Standard Model, and established a deep theoretical connection between quark-flavor mixing
and \C\Par~violation, two subjects that had previously been considered to be unrelated.
But this came at a cost: you need to have six-quarks. In 1971, thanks to Kiyoshi Niu,
this was not such a big stretch for Kobayashi and Maskawa, who were among the fortunate few who were already
convinced that there were (at least) four quarks. On the other hand, most of the world-wide particle physics
community outside of Japan was quite satisfied with three quarks.

\subsubsection{The first proposal for (and the naming of) charm}
\noindent
The distinction between electron- and muon-neutrinos had been established in 1962~\cite{Danby:1962nd},
and, in what was at that time the beginnings of the Standard Model, the electron and muon and their neutrinos
occupied two weak-isospin=1/2 spinors. In a 1964 paper, Bjorken and Glashow~\cite{Bjorken:1964gz} discussed an
expansion of $SU(3)$ to $SU(4)$ with the addition of another strangeness-like flavor quantum number. They formulated
their model in terms of the Sakata model, the predecessor of the quark model that used the proton-neutron isospin
doublet and isoscalar Lambda as basic constituents, with the Lambda replaced by a doublet that had a fourth baryonic
constituent with a non-zero value of the new quantum number. When reformulated in the context of the quark model,
which had just emerged at that time with three quarks, this was equivalent to adding a fourth quark with matching
patterns for the leptons and  quarks:
\begin{eqnarray}
  {\rm leptons}~~~~~~~~~~~~&~~&~~~~~~~~~~~{\rm quarks}\\
  \nonumber
  \begin{pmatrix} e^-\\ \nu_e \end{pmatrix}~\begin{pmatrix} \mu^-\\ \nu_\mu\end{pmatrix}~~~~~&~~&~~~~~~
    \begin{pmatrix} u\\ d \end{pmatrix}~\begin{pmatrix} \boldsymbol{c}\\ s \end{pmatrix}.
\end{eqnarray}
While their proposal was not very different from schemes that other authors had proposed around that time (see,
for example refs.~\cite{Tarjanne:1963zza,Maki:1963ye,Hara:1963gw}), Bjorken and Glashow gave the new flavor the
charismatic name ``charm,'' and that's the one that stuck; the fourth quark was known as the ``charmed quark''
(and not the grammatically awkward ``charm quark'') even before it was discovered. In retrospect, this
four-quark scheme seems like a pretty sound---almost obvious---idea;\footnote{In a passing comment in
  his original paper on quarks~\cite{GellMann:1964nj}, Gell-Mann mentioned a four-quark
  scheme that was ``parallel with the leptons'' as an interesting possibility.}
how could anyone dismiss such a simple and sensible suggestion? Nevertheless, for the next six years this idea
didn't go very far. By the time the GIM paper~\cite{Glashow:1970gm} appeared in 1970, the Bjorken-Glashow paper had
received a grand total of six citations. For some reason, there seem to have been a special affinity in the physics
community for the number three and an aversion to the number four.\footnote{In east Asian cultures the number four is
  considered to be very unlucky because the Chinese pronunciations of  their words for
  ``four'' and ``death'' are homophonous. As far as I know there is no such taboo in
  Western cultures.} 
Moreover, even the GIM paper that proposed a four-quark scenario as a compelling explanation for the suppression
of flavor-changing neutral-currents, an important theoretical issue at that time, didn't experience a boom
in citations until after the $\jpsi$ discovery, which finally convinced the world-wide physics community
that there were (at least) four quarks.

In 1975, soon after the $\jpsi$ discovery, Maiani (the ``M'' in GIM), who, at that time, was unaware
of the KM result,\footnote{See footnote~2 in Maiani's paper.}
wrote an interesting paper~\cite{Maiani:1975in} that contained all of the KM model and then some. But, thanks
to their three-year-long head start, it was Kobayashi and Maskawa---and not Maiani---that got to go to Sweden in
December 2008.

The six-quark era started in earnest in 1977, after the discovery of the $\tau$-lepton~\cite{Perl:1975bf}
and the $b$-quark~\cite{Herb:1977ek}, and the KM model was elevated to role  of being the most likely
mechanism for \C\Par~violation. Although the $b$-quark's charge=2/3\,partner, the $t$-quark, wasn't
discovered until 1995~\cite{Abe:1995hr,D0:1995jca}, there was very little doubt about its
existence.\footnote{There were some papers on ``topless models,''~\cite{Barger:1981vg,Branco:1978mr,Gorn:1980cj}
   including one with the  title: ``Does bare-bottom rule out the
  topless E6 model?''~\cite{Sarkar:1983yx}, but these did not get much attention.}
The only question was its mass; based on the mass pattern of the known quarks, the general consensus
was that it was probably in the $\sim$25--35~GeV range~\cite{Yamanaka:1981pa}. 

\subsection{``Discovery'' of the KM paper}
\noindent
As mentioned above, for over two years the KM paper remained completely unnoticed outside of Japan (and only
received limited attention inside Japan). It was finally brought to the attention of the world-wide physics
community in a curious set of circumstances that are recounted here.

But first it should be noted that version of the quark-mixing matrix that appeared in the KM paper and
is reproduced above in eqn.~\ref{eqn:KM-matrix-orig} contains a pretty obvious  typographical (or transcription)
error. Since the KM matrix nominally describes a rotation, if all three mixing angles and the \C\Par~phase are set
to zero, it should revert to the identity matrix, {\it i.e.},
\begin{equation}
  \label{eqn:V_CKM-limit}
  \boldsymbol{V}\xrightarrow{~(\theta_{1,2,3},\delta)\rt 0~}\boldsymbol{\mathcal I}
  =\begin{pmatrix} 1 & 0 & 0\\ 0 & 1 & 0 \\ 0 & 0 & 1 \end{pmatrix}
\end{equation}
However in the matrix that appears in the KM paper, the zero-angle limit for the lower-right diagonal element
$V_{tb}$, which should be $V_{tb}$$\rt$1, is incorrect:
\begin{equation}
  \label{eqn:KM-Vub-orig}
  V^{\rm KM}_{tb}=\cos\theta_1\sin\theta_2\sin\theta_3 -\cos\theta_2\sin\theta_3 e^{i\delta}
  \xrightarrow{~(\theta_{1,2,3},\delta)\rt 0~} 0.
\end{equation}

\begin{figure}[h!]
\centering
\includegraphics[width=0.80\textwidth]{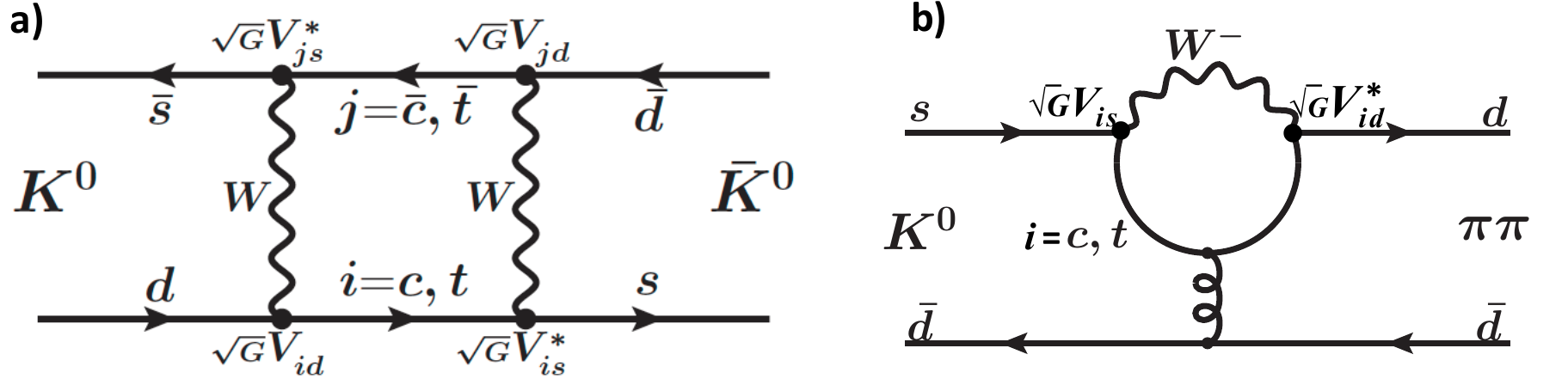}
\caption{\footnotesize {\bf a)} The $W$-exchange Standard Model box diagram  for $\Kz$-$\Kzbar$ mixing
 (not shown is the one with heavy quark exchange). In the KM six-quark model, the kaon's mass-matrix
  \C\Par-violating  parameter $\eps$ is produced by interference between the $c$-~and~$t$-quark contributions.
                    {\bf b)} The penguin diagram for direct-\C\Par-violating $\Ktwo$$\rt$$\pi\pi$ decays.
  }
 \label{fig:k-mix_ct-quarks-KM}
\end{figure}

The first paper to establish that the KM model could account for all that was known about \C\Par~violations at that
time, was by Pakvasa and Sugawara and published in the July~1976 issue of Physical Review~D~\cite{Pakvasa:1975ti}.
In their paper, they pointed out that in the six-quark model, a non-zero value of $\eps$, the neutral $K$-meson
mass-matrix \C\Par~violation parameter, of the correct magnitude would be produced by
the interference between the virtual $c$-~and~$t$-quark contributions to the $\Kz$-$\Kzbar$ mixing box-diagram
shown in Fig.~\ref{fig:k-mix_ct-quarks-KM}a.
They  also pointed out that in the KM picture, the penguin diagram\footnote{This was before
         penguin diagrams were called penguin diagrams.}
shown in Fig.~\ref{fig:k-mix_ct-quarks-KM}b, that would mediate direct-\C\Par~violating $\Ktwo$$\rt$$\pipi$ decays,
{\it i.e.}, the $\epsp$ parameter,
would be small and consistent with the then existing experimental limits. Since PRD is a widely
distributed physics journal, this paper provided the first awareness of the KM paper to the international particle
physics community.

In their paper,
Pakvasa and Sugawara  made no mention of the typo in the KM paper and included an expression for the  matrix
that they attributed to KM but, in fact, was different. This paper also had a typo that mistakenly
identified Toshihide Maskawa as K.~Maskawa in their citation to the KM paper.
Interestingly, many of the papers that immediately followed the Pakvasa-Sugawara paper used the
Pakvasa-Sugawara version of the CKM matrix with no mention of the mistake, and with citations to the KM paper
that had T. Maskawa incorrectly listed as K.~Maskawa (see, {it e.g.},
refs.~\cite{Lee:1976nx,Ellis:1977uk,Weinberg:1977maa,Altarelli:1977zq}).
This recurrence of the typo in the citations provided
pretty clear evidence that the Pakvasa-Sugawara PRD article was the source that researchers used for both
the matrix and the citation, and that the PTP paper itself was probably not very widely read.\footnote{But
  why didn't Pakvasa and Sugawara alert their readers about the problem with the matrix in the KM
  paper?  Sugawara's recollection is that when he first learned about the KM paper at a University
  of Tokyo physics seminar, he was impressed by their six-quark scheme and ``reconstructed it
  in [his] own way, without reading their paper carefully.'' When he and Pakvasa subsequently
  did their analysis and wrote up their results, they included Sugawara's version of the matrix,
  which was the KM version without the error, in their paper. In fact, Pakvasa and Sugawara
  remained unaware of the KM paper's typo. According to Sugawara, ``I never realized that
  the paper had this typo until it was recently pointed out to me.''}

But, in addition to the typo in their citation to the KM paper, the Pakvasa-Sugawara version of the KM
matrix had a problem of its own. In their paper, the KM expression for $V_{tb}$, given above in
eqn.~\ref{eqn:KM-Vub-orig}, was replaced by
\begin{equation}
  \label{eqn:KM-Vub-PS}
  V^{\rm PS}_{tb}=\cos\theta_1\sin\theta_2\sin\theta_3 -\cos\theta_2\cos\theta_3 e^{i\delta},
\end{equation}
which doesn't go to zero in the limit or zero mixing angles.  But neither does it go to 1,
instead,
\begin{equation}
  V^{\rm PS}_{tb}\xrightarrow{~(\theta_{1,2,3},\delta)\rt 0~} -1,
\end{equation}
and for $\delta=0$, has a negative determinant. Nevertheless, this version of the matrix is unitary, which is
the only essential requirement for a quark-flavor mixing matrix, and this form was used in much (but not all)
of the literature until 1984, when the currently widely accepted Chau-Keung parameterization was proposed and
soon thereafter endorsed by the Particle Data Group.

\section{Reparameterizing the CKM matrix}

\noindent
A rotation in three dimensions can be accomplished by three successive rotations: first by an angle $\theta_1$
around the $z$ axis that mixes $x$~and~$y$ [$(x,y)$$\rt$$(x^{\prime},y^{\prime})$)], then by $\theta_2$ about the
new $x^{\prime}$ axis that mixes $y^{\prime}$ and $z$ [$(y^{\prime},z)$$\rt$$(y^{\prime\prime},z^{\prime})$] and,
finally, by $\theta_3$ around the $y^{\prime\prime}$ axis that mixes $x^{\prime}$ and $x^{\prime}$
[$(x^{\prime},z^{\prime})$$\rt$$(x^{\prime\prime},z^{\prime\prime})$]. This is just one of of many ways that can be
used to specify a given rotation. For example, the order of the three rotations can be changed, and there is
freedom in the choice of the axes that are used to define the rotations. For these  different choices, the
values of the individual rotation angles are different, as are the expressions for each matrix element in
terms of these rotation angles. Ultimately, however, the numerical value of the magnitude of each matrix
element for any of these choices has to be the same, and independent of the choice of the individual rotations.
In addition, in the case of the CKM matrix, which is complex, rephasing invariance provides five independent
arbitrary phase parameters that can be attached to the various matrix  elements to establish whatever phase
convention may seem convenient. The physics content is independent of these parameterization choices.

On what basis should a parameterization be selected? In answer to this, Haim Harari suggested some
criteria for what he would consider to be a ``good'' parameterization. These included~\cite{Harari:1986xf}:
 \begin{itemize}
 \item There should be a simple relation between the most directly measurable matrix elements $V_{ij}$ and
   the quark mixing angles.
 \item The matrix elements above the diagonal, which correspond to kinematically allowed decay processes
   that are directly measurable, should have the simplest possible expressions.
 \item If possible, the \C\Par~violating phase should be linked to only one angle, and preferably the
   sine of that angle.
 \end{itemize}

 During the years immediately following the wide recognition of the KM paper, there was considerable effort aimed
 at finding a suitable parameterization. This was aided by concurrent experimental measurements of the relative
 magnitudes of the $V_{cb}$ and $V_{ub}$ matrix elements using $B$-mesons that were produced via $\ee$
annihilations at two colliders that existed at that time: PEP at SLAC  and CESR at Cornell. 

 \subsection{Experimental information about quark transitions}
\noindent
 The PEP project was initially conceived as a two  ring  proton-electron-positron collider with  an
 electron-positron ring that could support $E_{\rm cm}$=\,30~GeV $\ee$ collisions primarily to search for the
 top-quark (if its mass was less than 15~GeV), and a second 150~GeV proton ring that could support $e^{\pm}p$
 collisions with $E_{\rm cm}$\,$\approx$\,100~GeV for measurements of deep inelastic scattering at high energies and
 $Q^2$ values. The CESR collider was conceived during 1974 as a follow-up to the Cornell fixed-target
 $e^-p$ scattering program, and proposed to the U.S. National Science Foundation in 1975, soon after the
 $\jpsi$ discovery, as an $E_{\rm cm}$$\approx$\,16~GeV $\ee$ collider primarily aimed at studies of charmed
 particles. The PEP and CESR projects were both well underway when the $b$-quark was discovered in 1977.
 Fortuitously, the initial CESR design energy could comfortably operate in the $E_{\rm cm}$=\,9$-$11~GeV
 range, and cover the three narrow $\Upsilon (1S,2S,3S)$ resonances and the threshold region
 for $\ee\rt B\bar{B}$ meson pair production,  that was expected to be around
 $\Ecm$=\,$10.5$~GeV.\footnote{The CESR facility spent its  life operating at $\Ecm$$\sim$10--11~GeV.}

 The two projects started running in 1979 and soon thereafter provided convincing experimental evidence
 for a strong hierarchy among the weak interaction mixing angles for the three quark generations. It was
 already well established that transitions {\it within} the first quark generation, {\it e.g.}, $u$$\rt$$d$ and
 {\it within} the second generation ($c$$\rt$$s$) were strongly favored over transitions {\it between} the first
 and second generations ($s$$\rt$$u$ and $c$$\rt$$d$), {\it i.e.}, the Cabibbo angle. Experiments at PEP found
 that transitions {\it between} the second and third generation ($c$$\rt$$b$) are more suppressed those between the
 first and second generations, and the CESR experiment determined that transitions between the first and third
 generations ($u$$\rt$$b$) are the least favored of all. 

\begin{description}
\item[1$^{\rm st}$$\leftrightarrow$~2$^{\rm nd}$ generation:]~~~The suppression of strangeness-changing decays
  that was noted by Cabibbo in 1963, led to the realization that the relative strengths of the
  $u$$\rt$$d$ and the strangeness-changing $u$$\rt$$s$ transitions are modulated by factors of $\cos\theta_C$  and
  $\sin\theta_C$, respectively, where $\theta_C$\,=\,13$^{\circ}$ is the Cabibbo angle. Thus the diagonal
  element $V_{ud}$\,=\,$0.974$ is nearly unity and almost five times larger that its adjacent entry
  $V_{us}$\,=\,0.225.

\item[2$^{\rm nd}$$\leftrightarrow$~3$^{\rm rd}$ generation:]~~~In 1983, the MAC experiment at the  PEP collider
  used $b$-flavored hadrons (mainly $B^{\pm}$ and $B^0$ mesons) produced via the $\ee$$\rt$$b\bar{b}$
  annihilation process (about 1/10$^{\rm th}$ of the total annihilation cross section) to determine the
  $b$-quark lifetime by  measuring the impact parameters of charged leptons from $B$$\rt$$X\mu\nu$ and
  $X e \nu$ inclusive semileptonic decays, where $X$ is a hadronic system~\cite{Fernandez:1983az}.  (The
  definition of the impact parameter is  indicated in Fig.~\ref{fig:Vcb-blifetime}a.) This was a  difficult
  measurement: the parent $B$-meson's energy and direction were not precisely known on an event-by-event basis;
  there was contamination from  semileptonic decays of charmed mesons; and the mean value of the impact
  parameter that was eventually determined ($\approx$170\,$\mu$m) was substantially smaller than the
  experimental resolution  ($\sim$600\,$\mu$m) as well as the horizontal size of the beam-beam interaction
  region ($\sim$400\,$\mu$m). Nonetheless, the measured impact parameter distributions for muons and electrons
  shown in the upper and lower panels of Fig.~\ref{fig:Vcb-blifetime}b, respectively, both had excesses at
  positive values and these translated into a $b$-quark lifetime of $\tau_b$\,=\,$1.8\pm 0.7$~ps. Since this was
  about a factor of four times longer than the lifetime of the much lighter $c$-quark, it was a big surprise
  at that time. As discussed below, $b$$\rt$$c$ transitions are the $b$-quark's dominant decay mechanism,
  and the measured lifetime could be used to determine a value for $|V_{cb}|$ using the expression
  \begin{equation}
  \label{eqn:Vcb-vs-lifetime}
    \Gamma_b=1/\tau_b
        =|V_{cb}|^2\frac{G_F^2m_b^5c^4}{192\pi^3\hbar^7}(2_{\rm quarks}\times 3_{\rm colors} + 3_{\rm leptons}),
  \end{equation}
  which is the (textbook) expression for the muon lifetime tailored to the $b$-quark
  mass, modified to include the  effect of $|V_{cb}|$ on the coupling strength, and multiplied by the number
  of accessible final states: two quark flavors, three quark colors, and three types of leptons, as illustrated
  in Fig.~\ref{fig:Vcb-blifetime}c. The result was $|V_{cb}|$\,$\approx$\,0.04, and about a factor of five smaller
  than $|V_{us}|$, a difference that is similar to (but not exactly the same as) the factor of five
  Cabbibo suppression\footnote{The PDG 2020~\cite{Zyla:2020zbs} value for the $B$ meson
    lifetime is $\tau_b$=\,$1.519\pm 0.004$~ps and that for the $V_{cb}$ matrix element is
    $|V_{bc}|$\,=\,$0.0410\pm 0.0014$, which are both within the error ranges of the MAC measurements.}
  between $V_{us}$ and $V_{ud}$. The MAC results were confirmed by the MarkII and DELCO experiments at
  PEP~\cite{Lockyer:1983ev,DELCO:1984jng} and the TASSO experiment~\cite{TASSO:1984ult} at PETRA, an
  $E_{\rm cm}$=$40$~GeV $\ee$ collider at the DESY laboratory in Hamburg.
  \begin{figure}[!tbp]
  \centering
     \includegraphics[width=1.0\textwidth]{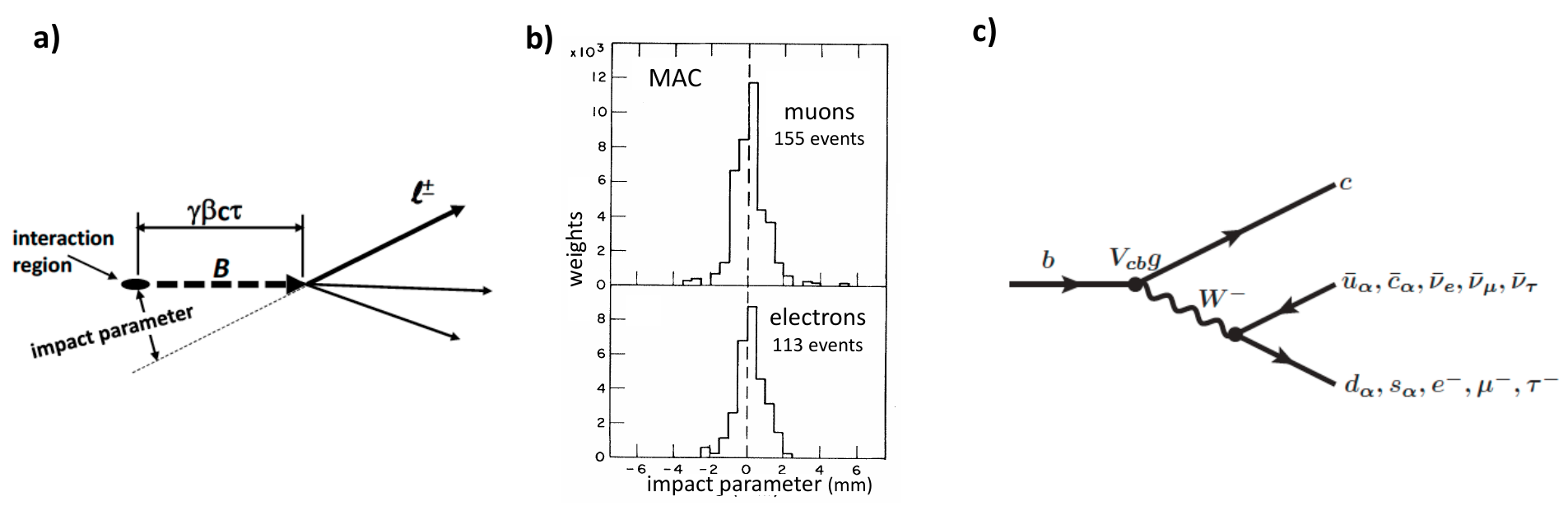}
     \caption{\footnotesize {\bf a)} A sketch of the projections of tracks from the semileptonic decay of a
       $B$ meson onto a plane perpendicular to the $\ee$ beamline, together with indications of the size of
       the beam-beam interaction region, and the definition of the impact parameter, {\it i.e.}, the distance
       of closest approach of the charged lepton track  to the $\ee$ interaction point. The unknown
       $B$ meson's direction was assigned with reasonably good accuracy to be along the event's thrust axis.
       {\bf b)} Impact parameter distributions for muons (upper) and electrons (lower) (from the MAC
       experiment~\cite{Fernandez:1983az}).  Here each entry is weighted by the inverse square of its
       measurement error. {\bf c)} A quark-line diagram for $b$-quark decays. The subscript $\alpha$ indicates
       the quark color.
       }
   \label{fig:Vcb-blifetime}
\end{figure}
\item[1$^{\rm st}$$\leftrightarrow$~3$^{\rm rd}$ generation:]~~~The CLEO experiment studied semileptonic
  $B$$\rt$$X\ell\nu$ decays of $B$ mesons that were produced at $E_{\rm cm}$\,=\,$10.58$~GeV, the
  peak\footnote{The $\Upsilon(4S)$ is the 3$^{\rm rd}$
      radial excitation of the $\Upsilon(1S)$, the lowest-lying  $J^{PC}$=\,$1^{--}$ ``bottomonium'' ($b\bar{b}$)
      meson, and the first that is above the $B\bar{B}$ ``open bottom'' threshold. At its peak, the cross
      section for $\ee$$\rt $$B\bar{B}$ is about 1~nb.}
  of the $\Upsilon(4S)$$\rt$$B\bar{B}$ resonance~\cite{CLEO:1984uti}.
  Since this energy is only 20~MeV above the $B\bar{B}$ threshold, the $B$ mesons are produced very nearly at
  rest (the boost factor is $\gamma\beta$\,=\,$0.062$) and the energy of a decay lepton ($\ell$\,=\,$e,\mu$) in the
  laboratory frame is very nearly equal to what it is in the $B$ meson rest frame. In $b$$\rt$\,$c\ell\nu$-mediated
  decays, the minimum mass of the hadronic system is $M^{\rm min}_X$=\,$m_{D}$\,=\,$1.86$~GeV, and this translates into a
  maximum lepton momentum of $p_{c\ell\nu}^{\rm max}$\,=\,$2.31$~GeV/$c$; in $b$$\rt$\,$u\ell\nu$-mediated decays, the
  mass of the hadronic system can be as light as $M^{\rm min}_X$=\,$m_{\pi}$, and the end-point momentum is
  $p_{u\ell\nu}^{\rm max}$\,=\,$2.60$~GeV/$c$.
  Thus, measurements of the lepton momentum spectra in the end-point region can be used to determine the
  relative strengths of the $b$$\rt$$c$ and $b$$\rt$$u$ transitions and extract the value of $|V_{ub}|^2/|V_{bc}|^2$.
  Figure~\ref{fig:Vub-spectrum} shows the measured momentum distributions for electrons ({\it upper}) and
  muons ({\it lower}) together with expectations for $b$$\rt$\,$c\ell\nu$ (dashed curves) and $b$$\rt$\,$u\ell\nu$
  (dotted curves). There are no events in the 2.31$<$$p_{\rm lepton}$$<$2.60~GeV/$c$ range that could be unambiguously
  attributed to $b$$\rt$$u\ell\nu$ decays and the shapes of the spectra are consistent with expectations for
  $\sim$100\% $b$$\rt$\,$c\ell\nu$ with no significant contribution from $b$$\rt$\,$u\ell\nu$. From these data,
  the CLEO group established a 90\%~CL upper limit\footnote{The PDG 2020~\cite{Zyla:2020zbs} value is
    $|V_{ub}|/|V_{cb}|$\,=\,$0.093\pm 0.004$.} of $|V_{ub}|/|V_{cb}|$$<$0.14.
  \begin{figure}[!tbp]
  \centering
     \includegraphics[height=0.4\textwidth,width=0.35\textwidth]{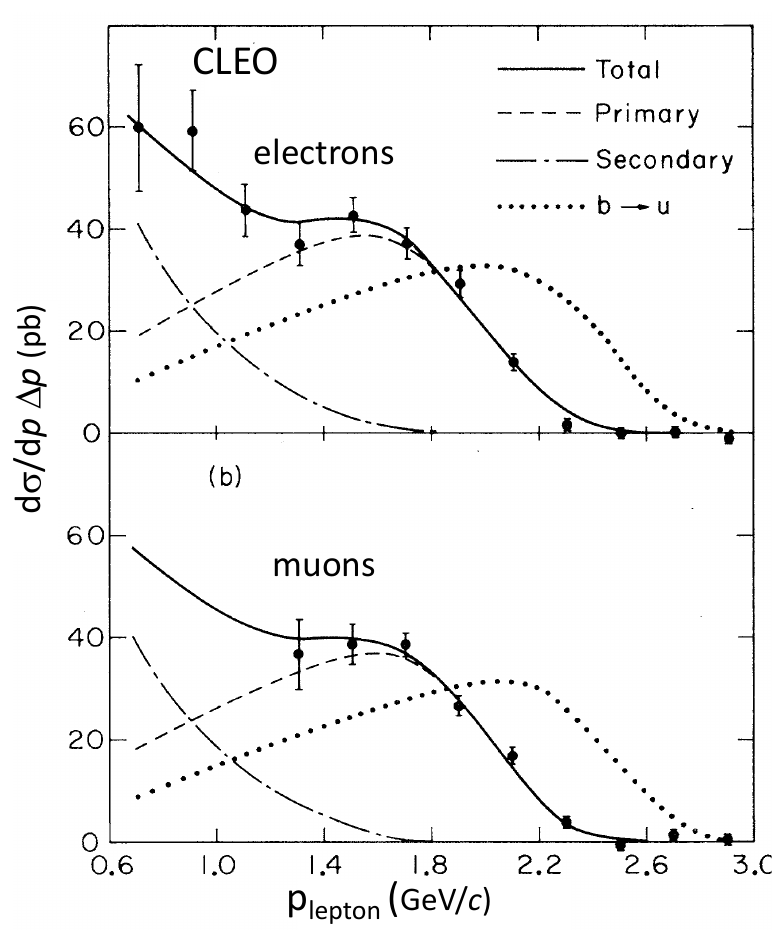}
     \caption{\footnotesize 
        Lepton momentum spectra for $B$$\rt$$X e\nu$ (upper) and $B$$\rt$$X\mu\nu$ (lower) decays.
        The dashed (dotted) curves are expectations for $b$$\rt$$c\ell\nu$ ($b$$\rt$$u\ell\nu$). The dash-dot
        curves shows the expectation for secondary leptons from semileptonic decays of $D$ decays produced
        via $B$$\rt$$D\ell\nu$  and of $\tau$ leptons from $B\rt X\tau\nu$. (From the CLEO
        experiment~\cite{CLEO:1984uti}.)  
        }
     \label{fig:Vub-spectrum}q
\end{figure}
\end{description}

\subsection{The Chau-Keung and Wolfenstein parameterizations}
\label{sec:Chau-Keung}
\noindent
Since the early parameterizations fail on all three of the Harari criteria, a better one was needed. Of a
number of proposed replacements\cite{Maiani:1975in,Chau:1984fp,Wolfenstein:1983yz,Gronau:1984dn,Harari:1986xf},
two continue to be widely used today: one by Chau and Keung~\cite{Chau:1984fp} and
the other by Wolfenstein~\cite{Wolfenstein:1983yz}.

\subsubsection{Chau-Keung parameterization}
\noindent
The parameterization proposed by Chau and Keung was specifically motivated by the occurrence of the
\C\Par~violating phase in the $V_{tb}=\cos\theta_1\cos\theta_2\cos_3-\cos\theta_2\cos\theta_3 e^{i\delta}$ term
in the Pakvasa-Sugawara version of the original KM parameterization that seemed to suggest that there is a
large \C\Par~violation that is confined to the ($t,b$) quark sector. This was in sharp contrast to the results
of detailed calculations of measurable \C\Par~violating effects that invariably resulted in small numbers that
involved factors of $\sin\theta_2\sin\theta_3 e^{i\delta}$.

(For clarity, in the following we follow the more transparent PDG notation that uses $\theta_{ij}$ to denote the
mixing angle around an axis that is perpendicular to the $(i,j)$ plane. The translation between $\theta_{1,2,3}$
and $\theta_{ij}$ is: $\theta_1$\,=\,$\theta_{12}$,~$\theta_2$\,=\,$\theta_{23}$,~{\rm and}~$\theta_3$\,=\,$\theta_{13}$.
In addition we abbreviate $\cos\theta_{ij}$~by~$c_{ij}$ and $\sin\theta_{ij}$ by $s_{ij}$.)
  
The Chau-Keung parameterization, which is exactly unitary, is given by the (right-to-left) sequence of
two-dimensional rotations:
  \begin{eqnarray}
    \label{eqn:CKM-Chau-Keung}
  V_{\rm CKM}&=&\overbrace{\begin{pmatrix}
  1   &   0     &    0     \\
  0   &  c_{23}  &   s_{23}  \\
  0   & -s_{23}  &   c_{23}
  \end{pmatrix}}^{\theta_{23}~{\rm about}~d''}
\overbrace{\begin{pmatrix}
    c_{13}          &   0    &  s_{13} e^{-i\delta} \\
      0            &   1    &  0               \\
      -s_{13}e^{i\delta~} &   0    &  c_{13}
\end{pmatrix}}^{\theta_{13}~{\rm about}~s'~({\rm incl.}~\delta )}
\overbrace{\begin{pmatrix}
   c_{12}  &  s_{12}  &  0  \\
  -s_{12}  &  c_{12}  &  0  \\
    0     &   0     &  1
\end{pmatrix}}^{\theta_{12}~{\rm about}~b}\\
\nonumber
  &=& \begin{pmatrix}
  c_{12}c_{13}                         &            s_{12}c_{13}                     &  s_{13}e^{-i\delta} \\
-s_{12}c_{23}-c_{12}s_{23}s_{13}e^{i\delta}~~&~~ c_{12}c_{23} - s_{12}s_{23}s_{13} e^{i\delta}~~&~~s_{23}c_{13}      \\
 s_{12}s_{23}-c_{12}c_{23}s_{13}e^{i\delta}~~&~~-c_{12}s_{23} - s_{12}c_{23}s_{13} e^{i\delta}~~&~~c_{23}c_{13}
\end{pmatrix},
  \end{eqnarray}
where it is apparent that this version passes the requirement that
${\boldsymbol V}$$\rt$$\boldsymbol{\mathcal I}$ when  $\theta_{ij}$$\rt$\,0.  Here Harari's three criteria
are satisfied: the $\theta_{12},\theta_{23},\theta_{13}$ mixing angles correspond to experimentally
distinct $u$$\lra$$s$,\,$c$$\lra$$b$,\,\&\,$u$$\lra$$b$ transitions, respectively; the \C\Par~phase factor is always
multiplied by a factor containing $s_{13}\,(=\,|\Vub|)$; and  the three above-diagonal terms have simple forms.
Moreover, in this parameterization, $\theta_{12}$, $\theta_{23}$ and $\theta_{13}$ are all in the first quadrant,
which means that $s_{ij}$ and $c_{ij}$ are all positive, the \C\Par~phase $\delta$ is also positive and in the
range $0\le\delta <2\pi$, and the matrix reduces to the $2\times 2$ Cabibbo matrix for
$\theta_{23}$\,=\,$\theta_{13}$\,=\,0. This has been the PDG parameterization of choice since
1986~\cite{ParticleDataGroup:1986kuw} and, according to recent measurements~\cite{Zyla:2020zbs},
\begin{eqnarray}
  \theta_{12}=13.09^{\circ}\pm 0.03^{\circ}~~~~~~~~~&~&~~~~~\theta_{23}=2.32^{\circ}\pm 0.04^{\circ}\\
  \nonumber
  \theta_{13}=0.207^{\circ} \pm 0.007^{\circ}~~~~~&~&~~~~~~\delta=68.53^{\circ}\pm 0.51^{\circ}.
\end{eqnarray}

\subsubsection{Wolfenstein parameterization}
\label{sec:wolfenstein-paramterization}
\noindent
The Wolfenstein parameterization is an approximation that employs a polynomial expansion in terms of
$\lambda$\,$\equiv$\,$\sin\theta_C$=\,0.2265 that reflects the hierarchical character of the CKM matrix. With an
accuracy up to ${\mathcal O}(\lambda^{3})$ it has the form:
\begin{equation}
  \label{eqn:wolfenstein-LO}
    V_{\rm CKM}=
    \begin{pmatrix}
      1-{\textstyle \frac{1}{2}}\lambda^2   &         \lambda &  A\lambda^3({\rho}-i{\eta}) \\
      -\lambda         & 1-{\textstyle \frac{1}{2}}\lambda^2  &   A\lambda^2  \\
      A\lambda^3(1-{\rho}-i{\eta})     &     -A\lambda^2           &          1
    \end{pmatrix}
    +{\mathcal O}(\lambda^4),
  \end{equation}
where the parameter $A$$\approx$\,0.8 accounts for the fact that the Cabbibo-like suppression between $\Vcb$ and
$\Vcd$ is about 20\% more severe than that between $\Vcd$ and $\Vud$.  In this parameterization, all of the
\C\Par~violation resides in the single parameter ${\eta}$ that is confined  to the $\Vtd$ and $\Vub$ corners
of the matrix where it is multiplied by  $\lambda^{3}A$, a small number.

This parameterization is very convenient and is widely used, but in a somewhat modified form that was suggested
by Buras and colleagues~\cite{Buras:1994ec}. In the Buras version, the Wolfenstein $\lambda$, $A$, $\rho$
and $\eta$ parameters are redefined in terms of the Chao-Keung angles to be exactly
\begin{eqnarray}
  \lambda&\equiv& s_{12} =\frac{|V_{us}|}{\sqrt{|V_{ud}|^2+|V_{us}|^2}}\\
  A\lambda^2&\equiv& s_{23} =\lambda\frac{|V_{cb}|}{|V_{us}|}\\
  A\lambda^3(\rho-i\eta)&\equiv&s_{13}e^{i\delta}=V^*_{ub}.
\end{eqnarray}
With these parameter definitions, $V_{ub}$ is the same as in the  Chau and Keung parameterizations,
and the higher corrections to $V_{us}$ and $V_{cb}$ start at ${\mathcal O}(\lambda^7)$ and
${\mathcal O}(\lambda^8)$, respectively.
In addition Buras~{\it et al.} defined new parameters $\bar{\rho}$ and $\bar{\eta}$ as
\begin{equation}
  \label{eqn:rho-bar-eta-bar-def}
  \bar{\rho}+i\bar{\eta}=-\frac{V_{ud}V^*_{ub}}{V_{cd}V^*_{cb}},
\end{equation}
in which case
\begin{equation}
  \bar{\rho}=\rho\big(1-\frac{\lambda^2}{2}\big)+{\mathcal O}(\lambda^4)~~~~{\rm and}
       ~~~~\bar{\eta}=\eta\big(1-\frac{\lambda^2}{2}\big)+{\mathcal O}(\lambda^4)
\end{equation}
and 
\begin{equation}
  V_{td}=A\lambda^3(1-\bar{\rho}+i\bar{\eta})+{\mathcal O}(\lambda^7).
\end{equation}

Although  the distinction between Wolfenstein's $\rho,\ \eta$  and (the nearly equal) $\bar{\rho},\ \bar{\eta}$
parameters may seem to be confusing and unnecessary, the latter parameters are preferred and are more generally
used (for reasons that are  discussed in detail in ref.~\cite{Charles:2004jd}).
The PDG review~\cite{Zyla:2020zbs} only provides values for the four (redefined) ``Wolfenstein parameters'' that,
in 2020, were
  \begin{eqnarray}
    \lambda=0.22650\pm 0.00048~~~~~&~&~~~~~A=0.790^{+0.017}_{-0.012}\\
    \nonumber
    \bar{\rho}=0.141^{+0.016}_{-0.017}~~~~~~~~~~~&~&~~~~~\bar{\eta}=0.357\pm 0.011.
  \end{eqnarray}
  \noindent

  The eqn.~\ref{eqn:wolfenstein-LO} form of the matrix is carefully tuned for \C\Par~violations in the $b$-quark
  sector that is produced by the imaginary parts of $V_{ub}$ and $V_{td}$, and, since all the matrix elements in the
  second row and column, {\it i.e.}, the ones that involve the strange and charmed quarks, are real, it
  is not applicable to descriptions of \C\Par~violation the $c$- and $s$-quark sectors. For this, Wolfenstein
  provided a version of the matrix that is expanded to include \C\Par-violating terms of ${\mathcal O}(\lambda^4)$
  in $V_{ts}$ and ${\mathcal O}(\lambda^5)$ in $V_{cd}$:
  \begin{equation}
    \label{eqn:wolfenstein-NLO}
    V_{\rm CKM}=
    \begin{pmatrix}
      1-{\textstyle \frac{1}{2}}\lambda^2 &  \lambda & A\lambda^3(\bar{\rho}- i\bar{\eta})) \\
              -\lambda(1+iA^2\lambda^4\bar{\eta}) & 1-{\textstyle \frac{1}{2}}\lambda^2  & A\lambda^2  \\
      A\lambda^3(1-\bar{\rho}-i\bar{\eta})     &     -A\lambda^2(1+i\lambda^2\bar{\eta})    &     1
    \end{pmatrix}
    +{\mathcal O}(\lambda^6).
  \end{equation}
  \noindent

  \noindent
  An expression involving the same parameters that is exactly  unitary is given by  Kobayashi
  in ref.~\cite{Kobayashi:1994ps}

\section{\C\Par~violation in $b$-quark decays?}

\noindent
In the KM model, \C\Par~violation in neutral $K$-meson decays, other than the $\eps$ mass-matrix parameter,
are mainly produced by complex phases in the upper-left
2$\times$2 corner of the KM matrix\footnote{$V_{ub}$ can contribute to kaon decays via
     penguin diagrams, but these contributions are suppressed by similar ${\mathcal O}(A\lambda^4)$ factors.}
that, in Wolfenstein's ${\mathcal O}(\lambda^5)$ parameterization,
(eqn.~\ref{eqn:wolfenstein-NLO}) is the confined to the phase of $V_{cd}$, and is tiny:
\begin{equation}
  \arg(V_{cd})=A^2\lambda^4\bar{\eta} = 5.9\times 10^{-4} \approx  0.03^\circ.
 \end{equation}
In contrast, the CKM matrix element for charmless decays of $B$-mesons that proceed via $b$\,$\rt$\,$u$ transitions
is $V_{ub}$, with a \C\Par~phase $\delta$ that (we now know) is $\delta$\,$\approx$\,$70^\circ$. If the kaon's
direct-\C\Par-parameter $\epsp$, caused  by  a tiny, fraction of a degree phase, can be measured in the 20$^{\rm th}$
century, the observation of \C\Par~violations produced by a~70$^\circ$ phase in $B$-meson decay in the 21$^{\rm st}$
century should be easy.

\vspace{1.5mm}
\noindent
{\it  Wrong!}~~~There are a number of important differences between the neutral kaon and $B$-meson systems
 that make the types of measurements that were used to discover and elucidate the properties of \C\Par~violations
 in the kaon system inapplicable to the  $B$-meson system.

\begin{description}
\item[{\it $B$-mesons have a huge number of different decay channels.}]~~\\
  In contrast to the $K$-meson system,
  where 99.98\% of $\KS$ decays are to either $\pipi$ or $\piz\piz$ final states, and 99.7\% of $\KL$ decays
  are to either $\pi\ell\nu$ or $\pi\pi\pi$ final states, $B$-mesons have hundreds of different decay modes almost
  all of which have, at best, fraction of a percent level branching ratios.
\item[{\it The $\Bz$-$\Bzbar$ mass eigenstates have very short, and nearly equal lifetimes.}]~~\
  In the
  $K$-meson system, the $\KS$ and $\KL$ mass eigenstates have large lifetime differences (0.1\,ns {\it vs.}
  52\,ns, respectively), and an essentially pure $\KL$ beam can be achieved by simply making a neutral beam
  line that is longer than several $\KS$ proper decay lengths. In comparison, the equivalent $\BH$ and $\BL$
  mass eigenstates have very nearly equal lifetimes of a mere 1.5~ps ($c\tau_B$\,=\,0.45\,mm), and there is no
  possibility for making a beamline of \C\Par-tagged $B$-mesons, much less one that distinguishes between the
  two different \C\Par~values.
\item[{\it $B$-meson decays to final states that are eigenstates of \C\Par~are infrequent.}]~~\\
  $\KS$ mesons
  decay almost exclusively to \C\Par-even $\pipi$ and $\piz\piz$ eigenstates; $\KL$ decays to \C\Par-odd
  $\pipi\piz$ and $\piz\piz\piz$ eigenstates occur with branching fractions of 19.5\% and 12.5\%,
  respectively. In contrast, the most prominent \C\Par-eigenstate decay mode for neutral $B$-mesons is
  $B^0$$\rt$$\jpsi \KS$, with a meager 0.045\% branching fraction.
\end{description}

\noindent
Moreover, as noted above, $V_{cb}$, which has no \C\Par-violating phase, has a magnitude that is an order of
magnitude larger than $V_{ub}$ and, thus, branching fractions for $b$\,$\rt$\,$u$ mediated ``non-charmed'' decays
of $B$ mesons are strongly suppressed. As a result the prospects for finding and studying \C\Par~violations
in the $B$-meson system looked pretty hopeless.

\subsection{Prospects for testing the KM \C\Par~mechanism: pre 1980}
\noindent
Sometime around 1979-80, Abraham Pais who, along with Murray
Gell-Mann was responsible for many of the fundamental theoretical  discoveries in the early
days of flavor physics, discussed the prospects  for \C\Par~measurements with charmed and beauty mesons
in a seminar at Rockefeller University in New York City, where he had this to say about \C\Par~violations
with heavy quarks~\cite{Sanda:2009zz}:
\begin{quotation}
  ``{ There is good news and bad news. The good news is that \C\Par~violation
         in a heavy meson system is quite similar to that of the $K$-meson
         system. The bad news is that there is little distinction like the $\KS$-$\KL$
         mass eigenstates. For heavy meson systems, both lifetimes are short}.''
\end{quotation}
In the audience was a young theorist Ichiro (Tony) Sanda, who recalls thinking at that time~\cite{Sanda:2009zz}:
\begin{quotation}
  ``{\it `\C\Par~violation in a heavy meson system is quite similar to that of the K meson system}?'---How
  could anything as interesting as \C\Par~violation be so uninteresting.''
\end{quotation}
and he resolved to find a way to prove that Pais was wrong.

\subsection{Tony Sanda's great idea}
\begin{figure}[!tbp]
\centering
\includegraphics[width=0.99\textwidth]{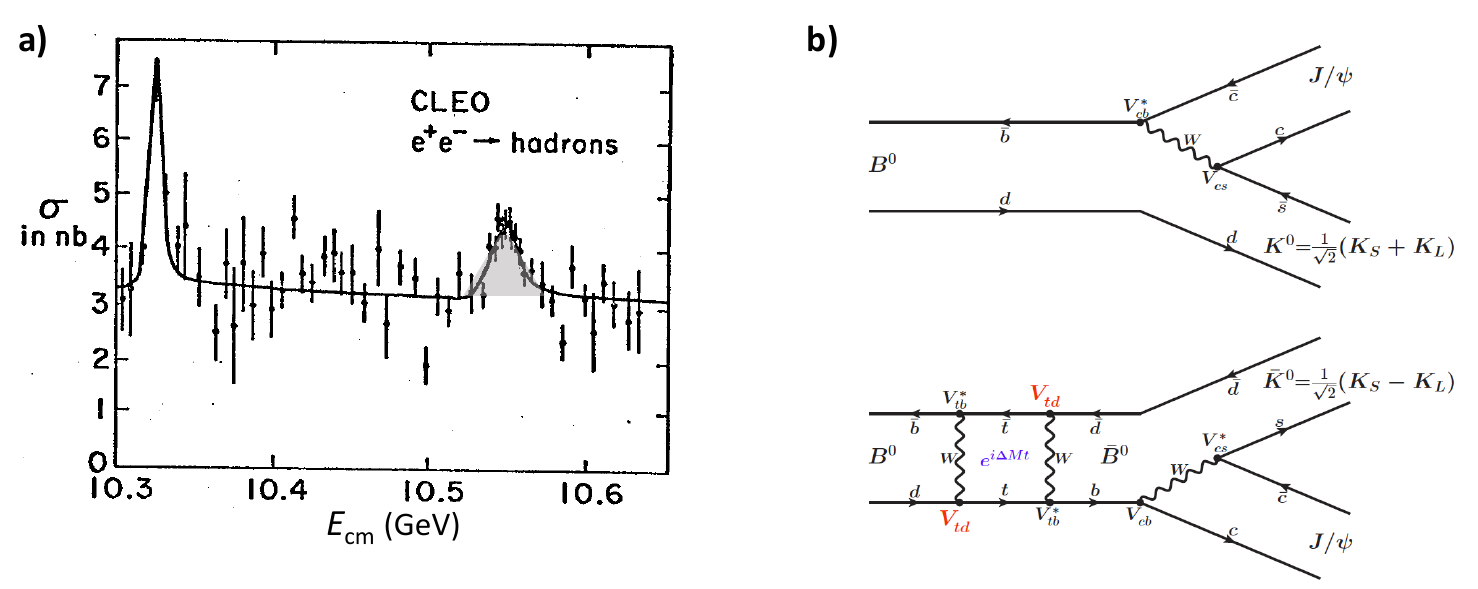}
\caption{\footnotesize 
  {\bf a)} CLEO's 1980 measurements of $\sigma(\ee$$\rt$${\rm hadrons})$ in the $\Upsilon(3S)~\&~\Upsilon(4S)$
  resonance region (from ref.~\cite{CLEO:1980tem}).
  {\bf b)}~Tree diagrams for $\Bz$$\rt$$\jpsi\Kz$;~$\Kz$$\rt$$\KS$~(or~$\KL$)
       ({\it upper}) and for $\Bzbar$$\rt$$\jpsi \Kzbar$;~$\Kzbar$$\rt$$\KS$~(or~$\KL$) after it
       is produced via $\Bz$$\rt$$\Bzbar$ mixing~({\it lower}).
       The $\Bzbar$ picks up a common phase $\delta_{\rm com}=\Delta M\tau$ from the mixing and a
       \C\Par~violating phase $\phi_{\CPa}$=$-2\beta$ from the two $V_{td}$  vertices in the mixing box.
}
 \label{fig:CLEO-1980-2box}
\end{figure}

\noindent
At this same time, I was one of the founding members of the CLEO experiment that was located on
the Cornell University campus in upstate New York, about a two-hour drive from New York City.\footnote{At
  that time I was at the University of Rochester, a two-hour drive in the opposite direction.}
The CESR $\ee$ collider was in its infancy and had a maximum instantaneous luminosity of
$\Lum$$\sim$\,$5\times 10^{30}$cm$^{-2}$s$^{-1}$.  We had just discovered the $\Upsilon(4S)$
resonance~\cite{CLEO:1980tem} and while running at $\Ecm$\,=\,10.58~GeV, its peak energy, we could collect
about 30~$\Upsilon(4S)$$\rt$$B\bar{B}$ events/day (see Fig.~\ref{fig:CLEO-1980-2box}a).  This was, at that time,
the most prolific source of $B$ mesons in the world and we were anxious to make good use of them. To this end,
my Cornell colleagues invited Sanda for some seminars. During his first seminar, the best strategy that Sanda had to
offer was a vague plan to search for a $\ell^+\ell^+$~{\it vs.}~$\ell^-\ell^-$ asymmetry in events of the type
\begin{equation}
\ee\rt\Bz\Bzbar\rt\ell^\pm\ell^\pm + {\rm anything},
\end{equation}
with the faint hope that somehow a measurable \C\Par~violation would show up. But this was very much like the
frequently performed $\Kz(\tau)$$\rt$$\pim\ell^+\nu$\,{\it vs.}\,$\Kz(\tau)$$\rt$$\pip\ell^-\bar{\nu}$ asymmetry
measurements, but without any of the above-listed advantages that make the neutral kaon system so special.
The experimenters in the audience, who were all hyped up to do great and wondrous things with the $B$ mesons
that they had worked so hard to produce, were noticeably disappointed. When Sanda got back to New York City,
he felt under strong pressure to come up with something that was new and unique to $B$~mesons.

A few months later, in his second seminar at Cornell, he did just that.  He proposed
a scheme that he developed in collaboration with Ashton Carter~\cite{Carter:1980hr,Carter:1980tk} for using
interference between the $\Bz$$\rt$$\KS\jpsi$~\&~$\Bz$$\rt$$\Bzbar$$\rt$$\KS\jpsi$ decay amplitudes that
eventually became the primary motivation for the BaBar and Belle asymmetric $B$-factory
experiments and was the basis for Kobayashi and Maskawa's Nobel prize.

The idea, which is illustrated in Fig.~\ref{fig:CLEO-1980-2box}\,b, was very elegant. You start with a
flavor-tagged $\Bz$ (or $\Bzbar$---here I use a tagged $\Bz$ to illustrate the idea) that can directly decay
via the $\Bz$$\rt$$\Kz\jpsi$ diagram show in the top panel, or decay via the indirect route where it first
mixes into a $\Bzbar$ that then decays via $\Bzbar$$\rt$$\Kzbar\jpsi$. But experiments don't distinguish
between $\Kz$ or $\Kzbar$ decays, instead they measure $\KS$ and $\KL$ decays. Thus, in events where a $\KS$ is
detected, the direct and indirect decay routes access identical final states and interfere. (Likewise
for events where a $\KL$is detected, except here the interference term has an opposite sign.)

The direct amplitude has no \C\Par~phase (at least not at leading order), but the indirect amplitude has an
extra factor of $V_{td}^{2}$  (not $|V_{td}|^2$!) and so, a \C\Par~phase of $-2\beta$. For tagged $\Bzbar$
decays, the CKM~factor is  $V_{td}^{*2}$ and the \C\Par~phase is $+2\beta$. Thus, the
$\Bzbar(\tau)$$\rt$$f_{\CPa}$ {\it vs.} $\Bz(\tau)$$\rt$$f_{\CPa}$ time-dependent asymmetry, where $f_{\CPa}$ is any
\C\Par-eigenstate is
\begin{equation}
  \label{eqn:B_2_CP-eigenstate-asymmetry}
  {\mathcal A}^{\CPa}_{B\rt f_{\CPa}}(\tau)=\frac{\bar{\Gamma}_{\Bzbar\rt f_{\CPa}}(\tau)-\Gamma_{\Bz\rt f_{\CPa}}(\tau)}
  {\bar{\Gamma}_{\Bzbar\rt f_{\CPa}}(\tau)+\Gamma_{\Bz\rt f_{\CPa}}(\tau)}
                                          = -\xi_{\CPa}\sin(2\beta)\sin(\Delta M_B\tau),
\end{equation}
where $\xi_{\CPa}$\,(=$-$1 for $\KS\jpsi$ and +1~for $\KL\jpsi$) is the \C\Par~eigenvalue of $f_{\CPa}$,
$\Delta M_B$\,=\, $M_H$-$M_L$ is the mass difference between the neutral $B_H$ and $B_L$ mass~eigenstates
({\it i.e.}, the $\Bz$-$\Bzbar$ mixing frequency), and $\tau$ is the proper time between the $\Bz$$\rt$$K\jpsi$
($\BCP$) decay and the flavor-specific  decay of the accompanying $\Bzbar$ meson ($\Btag$), whose decay products
are used to tag the $\Bz$ meson's flavor.  Note that in $\ee$~colliders, $\tau$ can be positive (when the
$\BCP$ decay occurs {after} the $\Btag$ decay), or negative (when the $\BCP$ decay occurs first),
and the time-integrated asymmetry is zero.

\subsubsection{Great idea! but is it practical?}
\noindent
The idea was new, and the mechanism was unique to $B$~mesons, but
there were many pieces that had to fall into just the right places for Sanda's proposal to have any chance
of being practical. Since at that time there was no experimental information about the $b$-quark-related CKM
elements or $\Bz$-$\Bzbar$ mixing, there was no way to form any opinion about the prospects for their favorability.

\begin{description}
\item[Tens of millions of tagged $B$$\rt$$f_{\CPa}$ decays would be required:]~~~\\
  In 1980, the world's best source of
  $B$-mesons was CESR, with a production rate of  $\sim$30~$B\bar{B}$ events/day, of which only half were the desired
  $\Bz\Bzbar$ pairs. Sanda's golden mode was $\Bz (\Bzbar)$$\rt$$\KS\jpsi$, which he estimated to have an
  ${\mathcal O}(10^{-3})$ branching fraction, and this implied that the fractional probability of usable events would
  be
\begin{equation}
  \label{eqn:golden-mode-effic}
  {\mathcal F}_{\KS\jpsi}<\underbrace{\BR (\Bz\rt\KS\jpsi)}_{\sim 10^{-3}}
  \underbrace{\BR(\KS\rt\pipi)\BR (\jpsi\rt\ell\ell)}_{\sim 10^{-1}}
  \underbrace{(\epsilon_{\rm trk})^4\epsilon^{\rm tag}_{\rm eff}}_{\sim 10^{-1}}\approx 10^{-5},
\end{equation}
where $\epsilon_{\rm trk}$ is the efficiency for charged track detection that, even in a nearly perfect detector,
cannot be much higher than $\epsilon_{\rm trk}$\,$\approx$\,0.85, and $\epsilon^{\rm tag}_{\rm eff}$\,$\approx$\,0.3
is an estimate of the maximum possible effective \Bty-flavor tagging efficiency. Thus, an ${\mathcal A}^{\CPa}_{{K\jpsi}}$
measurement with even modest precision would require $\sim$30\,M~$B\bar{B}$ events (and a million days of operation at 1980
state of the art collider and detector  performance levels).

\item[$\Bz$-$\Bzbar$ mixing had to be substantial, $|V_{cb}|$ had to be small, and $|V_{ub}|$ even smaller:]~\\
An essential part of the Sanda-Carter scheme is that the fraction
of $\Bz$ mesons that oscillate into a $\Bzbar$ before they decay has to be reasonably large. This meant
that $\Delta M_B$ would have to be similar in magnitude to $\Gamma_B$\,=\,$1/\tau_B$, which was then known to be
$\Gamma_B$$\approx$\,$4.4\times 10^{-10}$\,MeV. In the 1980s, when the $t$-quark mass was (almost
universally) expected to be $m_t$\,$\sim$\,35~\,GeV, calculations~\cite{Buras:1984pq} found
$\Delta M_B$\,$\approx$\,$1.2\times 10^{-10}$\, MeV. If this were the case, only $\sim$6\% of the tagged
$\Bz$-mesons would oscillate into a $\Bzbar$ before decaying. (After three lifetimes, the $\sin\Delta M_B\tau$
factor in eqn.~\ref{eqn:B_2_CP-eigenstate-asymmetry} would have barely reached 0.5.) In addition the
$B$ lifetime had to be relatively long: {\it i.e.}, $|V_{cb}|$$<$0.1, and $|V_{ub}|$$<$$|V_{cb}|$.

\item[The time sequence between the tag- and $f_{\CPa}$-decay has to be distinguished:]~~~~\\
  The asymmetry in eqn.~\ref{eqn:B_2_CP-eigenstate-asymmetry} has opposite signs for negative and positive
  values of $\tau$, which makes it essential to distinguish between events in which the $\BCP$ decays
  occurred first from those when it decayed last. The $B$ mesons that are produced in
  $\Upsilon(4S)$$\rt$$B\bar{B}$ decays have c.m.~momenta $p_B$\,=\,327\,MeV/$c$, corresponding to
  $\gamma\beta$\,=\,$0.062$ and have a mean decay distance of $\beta\gamma c\tau_B$\,=\,28$\mu$m,
  which is unmeasurably small in a c.m.~$\ee$~collider environment.  For a collider operating at the
  $\Ups(4S)$, it would be impossible to distinguish the time sequence of the $\BCP$ and $\Btag$ decays.

\end{description}

\noindent
Since the existence of six-quarks was pretty well established, the KM-mechanism provided a
compelling and almost obvious mechanism for explaining the existence of \C\Par~violation. However the
prospects or a conclusive experimental test of this idea seemed hopeless. The fortuitous set of
circumstances that made studies of \C\Par~violation in the neutral kaon system possible seemed unlikely
to be repeated.

\subsection{Three miracles}
\noindent
Nevertheless, in spite of these obstacles, Sanda maintained a nearly mystical belief that ``{\it Mother
Nature has gone out of Her way to show us \C\Par~violation, and She will also show us the way to the
fundamental theory}''~\cite{Sanda:2009zz}, and forcefully advocated an aggressive program of experimental
investigations of \C\Par~violation in the decays of $B$~mesons.  However, for the reasons itemized above,
Sanda's advocacy was initially met with considerable skepticism from his colleagues in both the theoretical
and experimental physics communities.\\
~\\
\noindent
And then three miracles occurred:
\begin{description}
\item[Miracle 1: ${\boldsymbol\Bz}$-${\boldsymbol\Bzbar}$ mixing was discovered at DESY:]~\\
  The most exciting event in flavor physics during the 1980s was the 1987
  discovery of a large signal for $\Bz$-$\Bzbar$ mixing by the ARGUS experiment at DESY~\cite{ARGUS:1987xtv}.
  The strength of the mixing was clear evidence the that top-quark mass, now known to be 173\,GeV, was nearly
  an order of magnitude larger than expected, which was shocking news to almost everyone. This discovery,
  coupled with the 1.5\,ps $B$-meson
  life-time measurements from PEP and PETRA that translated into $|\Vcb|$$\approx$\,0.04, and the suppression
  of $b$$\rt$$u$ relative to $b$$\rt$$c$ transitions meant that $|\Vub|$ was about a factor of ten smaller
  than $|\Vcb|$. These measurements confirmed Sanda's strong belief that Mother Nature would indeed help us
  ``find the way to the fundamental theory.''
  
\item[Miracle 2: Three-order-of-magnitude improvement in ${\boldsymbol \ee}$ collider luminosity:]~\\
  Advances in the understanding and modeling of beam dynamics, the use of separate magnet rings
  that enabled multibunch collisions, and major advances in RF feedback systems  provided the
  huge increases in the $\ee$$\rt$$\Upsilon(4S)$ production rate that were required by the
  experiment~\cite{Funakoshi:2014yfa,Seeman:2009zza}.
  
\item[Miracle 3: The invention of asymmetric $\ee$ colliders and innovations in detector technology:]~\\
  Pierre Oddone realized that an $\ee$ collider operating at the $\Upsilon(4S)$ resonance  with a modest
  ({\it i.e.}, factor of $\sim$2) difference between the $e^+$ and $e^-$ beam energies would produce boosted
    $B$-mesons with ${\mathcal O}(100\,\mu{\rm m})$ separation distances between the $\Btag$ and $\BCP$
    vertices~\cite{Oddone:1987up}. This idea, coupled with the concurrent development of high resolution
    silicon-strip vertex detectors that could measure such small displacements in a collider
    environment~\cite{Litke:1987xy,Batignani:1990jv,BaBar:1999vnr,Belle:2000mhy}, offered a realistic solution
    to the decay-time-sequence determination problem. Parallel improvements in detector performance levels in
    areas of particle identification~\cite{DIRC:1999mex,Iijima:2000uv}, and $\gamma$-ray~\&~$\KL$
    detection~\cite{BaBar:2001yhh,Belle:2000cnh,Belle:2000jde} advanced the state of the art levels of detection
  efficiencies and \Bty-flavor-tagging quality. 
\end{description}

\section{First measurements of the KM angle $\beta$ }
  \label{sec:First-Results}
\noindent
At leading order, measurements of the eqn.~\ref{eqn:B_2_CP-eigenstate-asymmetry} Carter-Sanda asymmetry
determines $\sin 2\beta$, where $\beta$\,=\,$\tan^{-1}\big(\bar{\eta}/(1-\bar{\rho})\big)$ and
$\bar{\eta}$~\&~$\bar{\rho}$ are the modified Wolfenstein parameters described in
Section~\ref{sec:wolfenstein-paramterization}. 
Prior to the summer of 2001, the best measurements of $\sin 2\beta$ were from CDF~\cite{CDF:1999ijp}
($0.79\pm 0.44$), BaBar~\cite{BaBar:2001ags} ($0.34\pm 0.21$)  and Belle~\cite{Belle:2001qdd} ($0.58\pm 0.34$).
The BaBar result was based on a sample of 23\,M $\Ups(4S)$$\rt$$B\bar{B}$ events and Belle, which was struggling with
electron cloud effects in the KEKB positron ring~\cite{Wang:2002ds}, had a smaller data sample of 11\,M
$\Ups(4S)$$\rt$$B\bar{B}$ events. Each of the three measurements were about $1.5\sigma$ from zero, and their weighted
average, $0.46\pm 0.17$ indicated a non-zero \C\Par~violation at the $\sim$2.5$\sigma$ level. The situation was
tantalizing, but not conclusive.

This changed in August 2001 when, in back-to-back articles in Physical Review Letters, BaBar, now with a sample
of 32\,M $B\bar{B}$ pairs, reported~\cite{BaBar:2001pki}
\begin{equation}
  \sin 2\beta= 0.59 \pm 0.14 \pm 0.05~~~~{\rm BaBar~(2001)},
\end{equation}
and Belle, with a 31\,M $B\bar{B}$ pair data sample, reported~\cite{Belle:2001zzw}
\begin{equation}
  \sin 2\beta= 0.99 \pm 0.14 \pm 0.06~~~~{\rm Belle~(2001)}.
\end{equation}

\begin{figure}[!tbp]
  \centering
     \includegraphics[height=0.4\textwidth,width=1.0\textwidth]{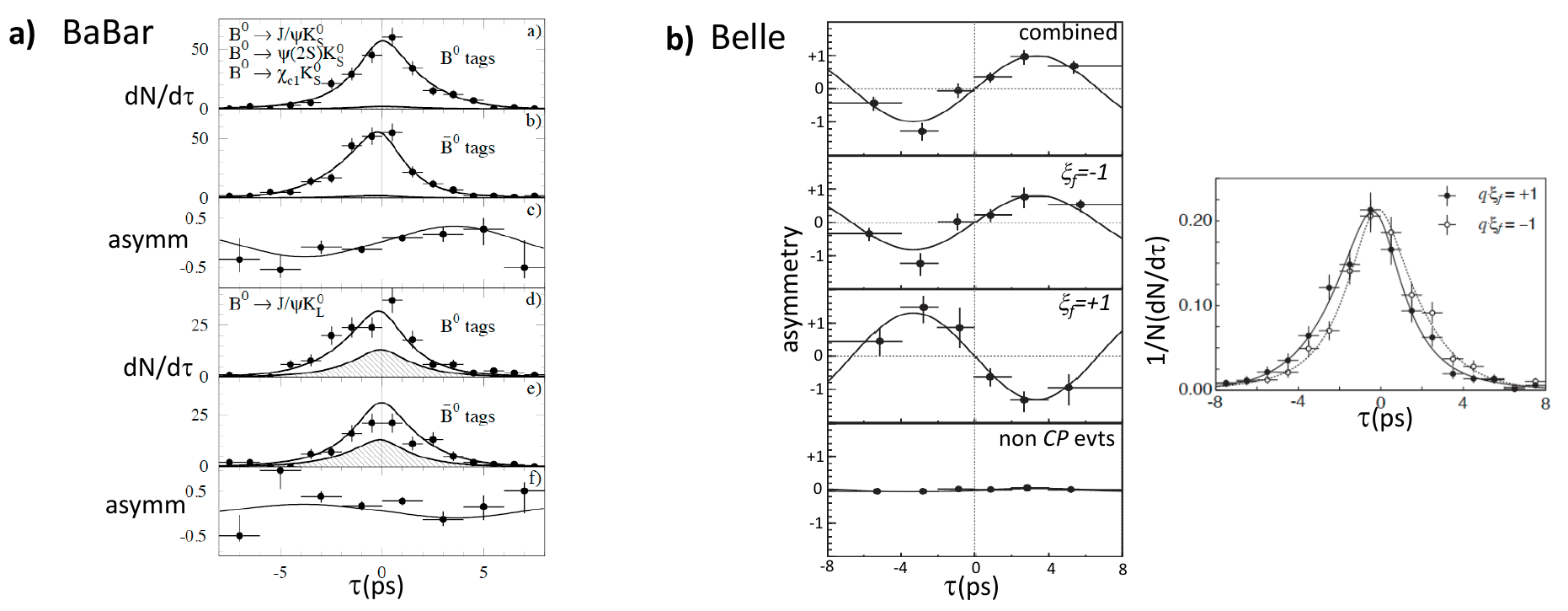}
     \caption{\footnotesize {\bf a)} BaBar results: the top three plots show the proper time
       distributions and fit results for $\Bz$-tags, $\Bzbar$-tags and their asymmetry for $\xi_{\CPa}$=$-1$
       $\BCP$ decays. The bottom three plots show the same distributions for $\xi_{\CPa}$=$+1$ $\BCP$ decays.
       The shaded areas indicate the background levels. (From ref.~\cite{BaBar:2001pki}). 
       {\bf b)} Belle results: the top three plots on the left show
       the time-dependent asymmetry and fit results for: (from top down) the combined
       ($\xi_{\CPa}$=$-1$ events {\it minus} $\xi_{\CPa}$=$+1$ events) samples; the $\xi_{\CPa}$=$-1$
       (mostly $\KS\jpsi$) events; and the $\xi_{\CPa}$=+1 ($\KL\jpsi$) events. The bottom plot shows results for
       non-\C\Par-eigenstate decay modes where no asymmetry is expected.
       The open circles and dashed curve on the right show background-subtracted time distributions
       for $q\xi_{\CPa}$=$-1$ events and the fit results. The solid circles and curve show the same quantities
       for $q\xi_{\CPa}$=$+1$ events.  (From ref.~\cite{Belle:2001zzw}).
    }
    \label{fig:LP2001}
\end{figure}

The BaBar data sample contained 803~$\BCP$ event candidates with a signal purity of about 80\%. The top three panels
in Fig.~\ref{fig:LP2001}a show the BaBar experiment's measured time distributions for $\xi_{\CPa}$=$+1$~$\BCP$ decays.
The uppermost plot shows the number of $\Bz$ tags where it is evident that there are more events with $\tau$$>$0 than
with $\tau$$<$0, while the $\Bzbar$ tags in the panel beneath it display an opposite pattern. The third panel
shows the bin-by-bin asymmetry where there is a clear indication of the sine-like behavior that is expected for
a \C\Par~violation as given in eqn.~\ref{eqn:B_2_CP-eigenstate-asymmetry}. The three lowest panels show the
corresponding results for $B\rt$$\KL$$\jpsi$ decays with $\xi_{\CPa}$=$-1$, where the $\tau$-dependent
asymmetries have opposite signs, again as expected.

Figure~\ref{fig:LP2001}b shows the results from the Belle experiment~\cite{Belle:2001zzw} that used
747~$\xi_{\CPa}$=$-1$~$B$-decay candidates (mostly $B$$\rt$$\KS\jpsi$) with 92\% purity and 569 $\xi_{\CPa}$=$+1$
$B$$\rt$$\KL\jpsi$ decay candidates with 61\% purity. The top plot on the left side of the panel shows the proper-time
distribution for the $\xi_{\CPa}$=$-1$ modes {\it minus} that for $\xi_{\CPa}$=$+1$ modes.
The 2$^{\rm nd}$\,and\,3$^{\rm rd}$ panels from the top show the $\xi_{\CPa}$=$-1$\,and\,$\xi_{\CPa}$=$+1$\,modes separately,
where the different \C\Par~modes have opposite-sign asymmetries as expected. The curves show the results of fits to the
data. The bottom panel shows the asymmetry for a large sample of self-tagged (non-\C\Par~eigenstate) $B$ decays
($\Bz$$\rt$$ D^{(*)-}\pip$, $D^{*-}\rho^+$, $K^{*0}(K^+\pim)\jpsi$, and $D^{*-}\ell^+\nu$), where a non-zero asymmetry
would have to be due to instrumental effects; the fit to this sample returned an asymmetry amplitude of $0.05\pm 0.04$.

The open circles in the plot on the right side of Fig.~\ref{fig:LP2001}b show the time distribution for $\Bzbar$-tags
($q$=1) in $\xi_{\CPa}$=$-1$ $\BCP$ decays plus $\Bz$-tags ($q$=-1) in $\xi_{\CPa}$=$+1$ decays ({\it i.e.},
$q\xi_{\CPa}$=$-1$), with the fit results shown as a dashed curve. The black dots and solid curve show the sum of the
opposite combinations ($q\xi_{\CPa}$=$+1$) and their fit result.

The BaBar and Belle results  excluded a zero value for $\sin 2\beta$ by $4\sigma$, and $6\sigma$, respectively,
and their combined significance established conclusively that \C\Par~symmetry is violated in the $B$-$\bar{B}$ mixing
process as as predicted by the KM six-quark model.  Makoto Kobayashi and Toshihide Maskawa shared
half of the 2008 Nobel Prize in Physics ``{\it for the discovery of the origin of the broken symmetry which
predicts the existence of at least three families of quarks in nature.}" The Nobel committee's remarks that
accompanied their announcement included the following:

\begin{quotation}
  ``[Kobayashi and Maskawa] explained broken symmetry within the framework of the Standard Model, but
    required that Model be extended to three families of quarks. These predicted, hypothetical new quarks have
    recently appeared in physics experiments. As late as 2001, the two particle detectors BaBar at Stanford,
    USA and Belle at Tsukuba, Japan, both detected broken symmetries independently of each other. The results
    were exactly as Kobayashi and Maskawa had predicted almost three decades earlier.''
\end{quotation}

During the two decades following the Belle and BaBar reports, there have been many hundreds of measurements of
non-zero \C\Par~violating symmetries in $B$ meson decays, mostly by BaBar and Belle, which continued operating
until 2008 and 2010, respectively. This program is being continued by the LHCb~\cite{Belyaev:2021cyr}, an
experiment specialized for heavy flavor physics at the CERN Large Hadron Collider that began operating
in 2010, and Belle~II~\cite{Belle-II:2010dht}, an upgraded version of Belle at KEK that began operating in 2018.
As is discussed in more detail in other contributions to this symposium, all of these hundreds of measured
\C\Par~violations are well explained as being due to the effects of the single KM \C\Par-phase angle $\delta$.\\

\vspace{1.5mm}
\noindent
{{The KM model has been  a remarkable success.}}

\section{A few comments on  mixing and \C\Par~violation in the neutrino sector}

\noindent
As mentioned above, the two-doublet nature of the of the leptons was identified in 1961~\cite{Danby:1962nd},
a decade before it was established for quarks.  The notion of neutrino-mixing was first proposed four years earlier
by Pontecorvo~\cite{Pontecorvo:1957qd} in 1957 and the PMNS neutrino mixing matrix was suggested Maki, Nakagawa
and Sakata~\cite{Maki:1962mu} in 1962, a year before Cabibbo's paper appeared. When the $\tau$-lepton was discovered
in 1975, the six-lepton picture was established and the PMNS matrix for neutrinos expanded to the same 3$\times$3
structure as the CKM matrix for quarks. If neutrinos are Dirac particles, the mathematics of the neutrino-flavor mixing
matrix are the same as the KM matrix with three mixing angles $\theta_{ij}$ and one \C\Par~violating phase
$\delta_{\CPa}$, the so-called the ``Dirac'' phase.  If neutrinos are Majorana particles, lepton number is not
conserved and the matrix's number of degrees of freedom increases by two and there are two additional \C\Par-violating
``Majorana'' phases~\cite{Schechter:1980gr} that have no measurable effects on neutrino mixing experiments (which
makes them hard to measure). The commonly used parameterization of the PMNS matrix that doesn't include the Majorana
phases uses same mixing angle definitions as the eqn.~\ref{eqn:CKM-Chau-Keung}) Chau-Keung version of the CKM matrix,
but with the sequence of rotations reversed, {\it i.e.},
\begin{equation}
  \label{eqn:PMNS-2-matrix-form}
  \begin{pmatrix}
    \nu_e\\
    \nu_{\mu}\\
    \nu_{\tau}\end{pmatrix}
 =\stackrel{\rm Atmospheric}{\begin{pmatrix}
    1 &   0   &   0  \\
    0 & c_{23} & s_{23}\\
    0 &-s_{23} & c_{23}\end{pmatrix}}
  \stackrel{\rm Reactor~experiments}{\begin{pmatrix}
   c_{13}          & ~0~  &e^{-i\delta_{\CPa}}s_{13}\\
    0             & ~1~  &   0   \\
  -e^{-i\delta_{\CPa}}s_{13} & ~0~  & c_{13}\end{pmatrix}}
  \stackrel{\rm ``Solar"}{\begin{pmatrix}
   c_{12} & s_{12} &  ~0~  \\
  -s_{12} & c_{12} &  ~0~  \\
    0    &       &  ~1~ \end{pmatrix}}
 \begin{pmatrix}
   \nu_1 \\
   \nu_2 \\
   \nu_3 \end{pmatrix},
 \end{equation}
where $\nu_1,\nu_2,\nu_3$ denote the three neutrino mass eigenstates and $\theta_{12}$, $\theta_{23}$,
and $\theta_{13}$ are the ``solar,'' ``atmospheric,'' and ``reactor'' neutrino mixing angles.
The explicit form of the matrix is:\footnote{\C\Par-violating Majorana phases $\psi_1$\,\&\,$\psi_2$ can be included by
  $U^{\rm Maj}_{\rm PMNS}$\,=\,$U^{\rm Dirac}_{\rm PMNS}{\mathbb P}$, where
  ${\mathbb P}$\,=\,$\begin{pmatrix} e^{i\psi_1} & 0 & 0\\ 0 & e^{i\psi_2} & 0\\0 & 0& 1\end{pmatrix}$.}
\begin{eqnarray}
    \label{eqn:PMNS-2-matrix-dirac}
  U^{\rm Dirac}_{\rm PMNS} &=& \begin{pmatrix}
  ~U_{e1}   &~U_{e2}   &~U_{e3}   \\ 
  ~U_{\mu 1} &~U_{\mu 2} &~U_{\mu 3} \\
    ~U_{\tau 1}&~U_{\tau 2} &~U_{\tau 3}\end{pmatrix}\\
  \nonumber
         &=& \begin{pmatrix}
         ~c_{12}c_{13}                      &       ~s_{12}c_{13}                &~s_{13}e^{-i\delta_{\CPa}}\\
 -s_{12}c_{23}-c_{12}s_{23}s_{13}e^{i\delta_{\CPa}} &~c_{12}c_{23}-s_{12}s_{23}s_{13}e^{i\delta_{\CPa}} &~s_{23}c_{13} \\ 
 s_{12}s_{23}-c_{12}c_{23}s_{13}e^{i\delta_{\CPa}} &-c_{12}s_{23}-s_{12}c_{23}s_{13}e^{i\delta_{\CPa}} &~c_{23}c_{13}
 \end{pmatrix}, 
\end{eqnarray}
and the PDG~2020~\cite{Zyla:2020zbs} world averages for the three rotation angles are:
\begin{eqnarray}
  \label{eqn:PMNS-angles}
  \sin^2\theta_{12} &=& 0.307\pm 0.013~~~~\Rightarrow \theta_{12}= 33.6^{\circ}\pm 0.8^{\circ}\\
  \nonumber
  \sin^2\theta_{23} &=& 0.545\pm 0.021~~~~\Rightarrow \theta_{23}= 47.6^{\circ}\pm 1.2^{\circ}\\
  \nonumber
  \sin^2\theta_{13} &=& 0.0218\pm 0.007~~\Rightarrow \theta_{13}= 8.48^{\circ}\pm 0.14^{\circ}.
\end{eqnarray}
\noindent
Although mixing in the quark- and lepton-sectors have the same mathematical structure, there are
important differences in their practical applications, including:

\begin{figure}[!tbp]
\centering
\includegraphics[width=0.99\textwidth]{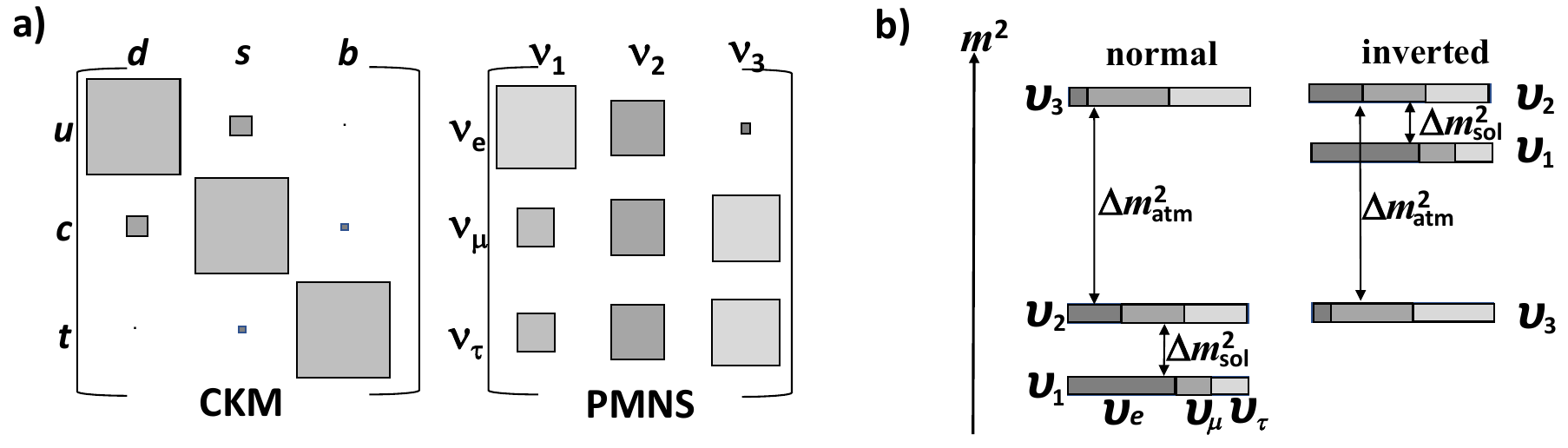}
\caption{\footnotesize {\bf a)} The area of the squares illustrates the magnitudes of
  the CKM ({\it left}) and PMNS ({\it right}) matrix elements.
  {\bf b)} A not-to-scale illustration of the normal and  ({\it left}) and inverted ({\it right})
  neutrino mass hierarchy.  The flavor content of each of the $\nu_1,\nu_2,\nu_3$ mass eigenstate
  is indicated by different shades of gray.
  }
 \label{fig:PMNS-matrix}
\end{figure}

\vspace{1.5mm}
\noindent
    {\it Strikingly different hierarchies:}~~~The  three PMNS mixing angles listed above differ in values
    by at most a factor of five, in contrast with the the CKM-matrix where the corresponding mixing
angles are smaller and span a two-order-of-magnitude range in magnitudes:
\begin{equation}
  \theta^{\rm CKM}_{12}\approx 13.0^{\circ}~~~~\theta^{\rm CKM}_{23}\approx 2.4^{\circ}
           ~~~~\theta^{\rm CKM}_{13}\approx 0.20^{\circ}.
\end{equation}
Differences between the two matrices are illustrated in Fig.~\ref{fig:PMNS-matrix}a, where
the areas of the squares are proportional to the magnitudes of the matrix elements.  Here Nature
has been helpful again;  if the PMNS matrix had a hierarchy that was similar to that of the CKM matrix,
neutrino oscillations and neutrino masses would likely not have been discovered.

\vspace{1.5mm}
\noindent
    {\it They operate in opposite ``directions:''}~~~In the CKM case, the quark mass-eigenstates are well known and
    the matrix is used to determine the flavor states. For the PMNS case, the neutrino flavor
    states are well known and the matrix is used to determine the mass eigenstates. Oscillation
    experiments have determined  the mass-difference hierarchies shown in Fig.~\ref{fig:PMNS-matrix}b,
    where~\cite{Zyla:2020zbs}
   \begin{equation}
  \nonumber
  {\rm atmos:}~\Delta m^2_{32} = \pm(2.44\pm 0.03)\times 10^{-3}{\rm (eV)^2}~~~~~~{\rm and}~~~~~~
      {\rm solar:}~\Delta m^2_{21} = (7.53\pm 0.18)\times 10^{-5}{\rm (eV)^2}.
\end{equation}
    The still unknown sign of $\Delta m^2_{32}$\,(=\,$m^2_{3}$-$m^2_2$) leaves two possible
    hierarchies as shown in Fig.~\ref{fig:PMNS-matrix}b.
    
\vspace{1.5mm}
\noindent
    {\it Quarks decay, neutrinos (probably) don't:}~~~CKM-related measurements are almost entirely based on
    measurements of decay processes with a formalism for oscillations and \C\Par~violations is done in
    the hadron's restframe.  In contrast case, PMNS-related measurement always involve production
    processes that produce pure flavor states and detection experiments located at some baseline distance that
    determine how the neutrino flavor-contents changed during their propagation to the detector.  Since the
    neutrino's restframe is ill defined,\footnote{If the lightest neutrino has zero mass
        it doesn't have a restframe.}
    the formalism is usually done in the laboratory frame and expressed in terms of $E_\nu$-\,and\,$\Delta m^2$-dependent
    oscillation lengths, {\it i.e.},
    \begin{equation}
      l(E_\nu)\equiv\lambda/2\pi = E_\nu/1.27\Delta m^2,
    \end{equation}
    where the factor of 1.27 is specific to ($l, E_\nu ,\Delta m^2$)
      units of (m,\,MeV,\,eV$^2$), or (km,\,GeV,\,eV$^2$).
    For the atmospheric and solar mass differences given above, these are   
    \begin{equation}
  \nonumber
  l^{\rm atm}(E_\nu)\approx 320~{\rm (km/GeV)}\times E_\nu~~~~~{\rm and}~~~~
                   l^{\rm sol}(E_\nu)\approx 10,500~{\rm (m/MeV)}\times E_\nu,
\end{equation} 
where the units km/GeV and m/MeV are equivalent (and interchangeable),  with km/GeV usually attached to
atmospheric and m/MeV to solar for historical reasons.

The 320~km atmospheric length is long, but not impossibly long. It is nearly the same as the 295\,km distance
between J-PARC and  Super Kamiokande (and the soon-to-be-completed HyperK detector), and corresponds to a 90$^\circ$
phase-change-induced oscillation maximum for
600~MeV muon-neutrinos that can be copiously produced by the J-PARC synchrotron. The 10.5~km solar length corresponds
to an oscillation maximum for $E_\nu$=\,3\,MeV reactor anti-electron-neutrinos at a baseline of 50\,km, which is
the baseline for the JUNO reactor neutrino experiment that will soon be operating in China. The
HyperK~\cite{Hyper-Kamiokande:2018ofw} and JUNO~\cite{JUNO:2015sjr} experiments will certainly not be easy, but they
will be done. If the neutrino oscillation lengths were factors of two or more longer, both experiments would be much more
difficult, if not impossible.

\section{Conclusions}

\noindent
The related subjects of \C\Par-violation and quark~\&~neutrino flavor mixing provide peeks into some of Nature's
most intimate secrets. When Fitch and collaborators measured the $\sim$0.2\% branching fraction for $\KL$$\rt$$\pipi$
in 1963, they were seeing the influence of the $t$-quark that wasn't discovered until thirty years later (see
Fig.~\ref{fig:k-mix_ct-quarks-KM}).  Thanks to Kobayashi and Maskawa, the existence and most of the properties
of the $t$-quark (other than its mass) were pretty well understood more than twenty years before it was discovered.

\subsection{Nature has been kind}

\noindent
(Einstein famously said ``The Lord God is subtle, but not malicious.'')

\noindent
A common thread that characterizes this entire story is that we have been able to probe these subjects at considerable
depth, even though it didn't {\it a priori} have to be that way.  The 0.2\% $\KL$$\rt$$\pipi$ branching fraction is as
large as it is because of the phase-space suppression of the partial decay width for the \C\Par-allowed
$\KL$$\rt$$\,3\pi$ mode that has a $Q$-value of only 83~MeV. This is a consequence of the relative masses of the
$K$-~and~$\pi$-mesons: if the kaon mass were higher and/or the pion mass were lighter, the $\KL$$\rt$\,$3\pi$ partial
width would be larger, and the $\KL$$\rt$$\pipi$ branching fraction would be reduced. It wouldn't take very large
changes in these masses to push the $\pipi$ branching fraction down to a value that was below the Fitch experiment's
sensitivity level. Unlike parity violation, which is a huge effect that is difficult to miss, the particle physics
consequences of \C\Par-violations in processes other than $\KL$ decays have only been seen in elaborate, highly
focused experiments that, absent the $\KL$ measurements, very likely wouldn't have occurred. 

As described above, the KM phase  was only measurable because Nature's parameters are aligned in a way that meet the
stringent requirements that were first identified in the Carter-Sanda papers, including  a strong
$|V_{us}|$$>$$|V_{cb}|$$>$$|V_{ub}|$ hierarchy and a large enough $t$-quark mass to produce a $B$-$\bar{B}$ mixing rate
that is close to the $B$-meson lifetime, but not so large that the mixing rate was much faster than the lifetime. As a
result, the phase could be measured, as could all three of the CKM matrix's mixing angles.

Remarkably, the same thread extends into the neutrino sector in which the PMNS matrix has a very different,
almost flat hierarchy that facilitated the discovery of neutrino oscillations. This hierarchy has also enabled precise
measurements of all three of its mixing angles and will likely allow for a determination of its Dirac phase in the
not-so-distant future. Moreover, and as mentioned above, Nature's choices of the differences between the
$\nu_1,\nu_2,\nu_3$ eigenstate masses have made these measurements realizable.

\subsection{What's next?... \C\Par\T?}
\noindent
Although \C\Par~violations only show up as small, subtle effects in particle physics experiments, the influence
of \C\Par~violations on the evolution of the Universe is glaringly obvious~\cite{Sakharov:1967dj}. However, the
\C\Par~violations that we measure in $K$-, $B$- and (recently~\cite{LHCb:2019hro}) $D$-meson decays that are
produced by the KM mechanism for quarks and, likely, leptons, cannot nearly account for the matter-antimatter asymmetry
of the Universe (see, {\it e.g.}, ref.~\cite{Canetti:2012zc}). There must be other sources of C\Par~violation that have
yet to be discovered, and these are the primary motivations for the LHCb and Belle II experiments. With the KM phase,
Nature has given us a  glimpse of \C\Par~violation, but not the whole story.

But what about \C\Par\T? The \C\Par\T~theorem~\cite{Schwinger:1951xk} states that any quantum field theory that is
{\it Lorentz invariant}, has {\it local point-like interaction vertices}, and is {\it Hermitian} ({\it i.e.}, conserves
probability) is invariant under the combined operations of \C,\,\Par\,and\,\T. Since the three QFTs that make up the
Standard Model---QED, QCD, and Electroweak theory---all satisfy these criteria, \C\Par\T~symmetry has been elevated to
a kind of  hallowed status in particle physics. But the non-renormalizability of quantm gravity calls into
question the validity of the locality assumption~\cite{Weinberg:1980kq}, and suggests that at some scale, \C\Par\T~has
to be violated. Strictly speaking, this violation only has to occur at impossibly high energies near the
${\mathcal O}(10^{19}\,{\rm GeV})$ Planck scale, but maybe the same thread of Nature that gives us a taste of
\C\Par~violations at energy scales well below the scale needed to explain the baryon asymmetry of the Universe,
will give us a hint of \C\Par\T~violation at scale below the one that's needed to rescue quantum gravity.  In any case,
since it is a fundamental feature of the Standard Model, \C\Par\T~invariance should be routinely challenged at
the highest feasible experimental sensitivities.

To date the most stringent test of the \C\Par\T~prediction that particle-antiparticle masses are equal comes from
kaon physics\footnote{The \C\Par\T~test in kaon physics involves a comparison the phase of the $\eta_{+-}$
  \C\Par~violation parameter in $\KL$$\rt$$\pipi$ decays with the ``superweak phase''
  $\phi_{\rm SW}$$\equiv$$\tan^{-1}\big(2(M_{\KL}$$-$$M_{\KS})/(\Gamma_{\KS}$$-$$\Gamma_{\KL})\big)$.}
~\cite{Schwingenheuer:1995uf,Apostolakis:1999zw} and sets a 90\% C.L.~limit on the $\Kz$-$\Kzbar$ mass
difference of
\begin{equation}
  |M_{\Kzbar}-M_{\Kz}|<5\times 10^{-19}~{\rm GeV},
\end{equation}
which is seven orders-of-magnitude more stringent than that for $m_{\bar{e}}$$-$$m_{e}$ and nine orders-of-magnitude
more stringent than the $m_{\bar{p}}$$-$$m_{p}$ limit.  This high sensitivity is because of the magic of the virtual
processes in the Fig.~\ref{fig:k-mix_ct-quarks-KM}a box diagram and the unique properties of the neutral kaons.

The kaon result is based on experiments done nearly thirty years ago with data samples containing tens of millions of
$K$$\rt$$\pipi$ decays.  One of the reasons these measurements have not been updated since then is that technologies for
improved sources of flavor-tagged neutral kaons have not been pursued. However, if the above-described three order of
magnitude improvements in $\ee$ collider luminosity that were developed for the $B$-factories would be applied to a
dedicated collider operating at $\jpsi$ mass peak, multi-billion-event/year samples of flavor-tagged neutral kaons,
produced via $\jpsi$$\rt$$K^\mp\pi^\pm\Kz(\Kzbar)$ decays, would be available to support \C\Par\T~tests
with more than an order-of-magnitude improved sensitivity. More modest improvements in \C\Par\T~sensitivity would be
provided by new colliders in the $\tau$-charm energy range that are being proposed in China~\cite{Achasov:2023gey} and
Russia~\cite{Barnyakov:2020vob}, if they spend sufficient time operating at the $\jpsi$ peak.

Maybe during the next sixty years, \C\Par\T~violation studies will prove to be as interesting and provocative as
\C\Par~studies have been during the past sixty years.
 
\section*{Acknowledgment}

\noindent
I thank the organizers of KM50 for inviting me to participate in this interesting symposium, the
editors of PTEP for inviting this manuscript, and this paper's referee, whomever he or she may be, for many
important corrections and helpful suggestions. This work was supported in part by the National Research Foundation of
Korea under Contract No. NRF-2022R1A2C1092335.

\let\doi\relax

\bibliographystyle{ptephy}    
\bibliography{KM50_solsen}      






\end{document}